\documentclass[floats,prb,twocolumn,superscriptaddress,preprintnumbers,amssymb]{revtex4}

\usepackage{graphicx}

\def\be{\begin{equation}}
\def\ee{\end{equation}}
\def\ba#1{\begin{array}{#1}}
\def\ea{\end{array}}
\def\bn{\begin{enumerate}}
\def\en{\end{enumerate}}

\def\r{\right}
\def\l{\left}

\def\summ{\sum\limits}

\def\c{\hat{c}}
\def\cd{\hat{c}^{\dagger}}

\def\H{{\cal H}}
\def\intt{\int\limits}
\def\prodd{\prod\limits}

\def\n{n_B}
\def\bmize{\begin{itemize}}
\def\emize{\end{itemize}}
\def\nbar{\overline{n}}

\begin{document}
\title{Superfluid-insulator transition in Fermi-Bose mixtures and the
orthogonality catastrophe}
\author{G. Refael}
\affiliation{Dept. of Physics, California Institute of
  Technology, 1200 E. California Blvd., Pasadena, CA 91125}
\author{E. Demler}
\affiliation{Department of Physics, Harvard University,
17 Oxford st., Cambridge, MA 02138}

\begin{abstract}

The superfluid-insulator transition of bosons is strongly modified by
the presence of Fermions. Through an imaginary-time path integral
approach, we derive the self-consistent mean-field transition line, and account for both the static and dynamic screening
effects of the Fermions. We
find that an effect akin to the fermionic orthogonality catastrophe,
arising from the fermionic screening fluctuations, suppresses
superfluidity. We analyze this effect for various mixture parameters and
temperatures, and consider possible signatures of the orthogonality
catastrophe effect in other measurables of the mixture.

\end{abstract}

\maketitle

\section{Introduction}

The superfluid to insulator (SI) transition of bosons provides a
conceptual framework for understanding quantum phase transitions in
many physical systems, including superconductor to insulator
transition in films\cite{GoldmanHaviland,HebardPaalanen,SteinerKapi,Shahar1,DynesG}, wires\cite{Lau,Bezryadin1,Bezryadin2,Chang}, Josephson junction
arrays\cite{Haviland1,Haviland2,Clarke}, quantum Hall plateau
transitions\cite{ZHK}, and magnetic
ordering\cite{AffleckM}. Theoretical work on this subject elucidated many
dramatic manifestations of the collective quantum behavior in both
equilibrium properties and out of equilibrium
dynamics\cite{FWGF,FisherFilms,SachdevBook}.
In many cases, however, we need to understand the SI transition not in
its pristine form, but in the presence of other degrees of freedom.
For example, in the context of the superconductor to insulator
transition in films and wires, there is often dissipation due to
fermionic quasi-particles, which may dramatically change the nature
of the transition\cite{Clarke,Kapi1,Kapi2,RDOF2,Fink3,VishMoore}. Remarkable progress achieved in recent
experiments with ultra-cold atoms in optical lattices (see
ref. \onlinecite{bloch-2007}  for a review) makes these
systems particularly well suited for examining quantum collective
phenomena, not only as exhibited directly in the superfluid phase, but also
through its interplay with other correlated systems under study.

A class of systems that can provide a new insight on the role of
dissipation and of a fermionic heat bath on the superfluid-insulator
transition are Bose-Fermi mixtures of ultra-cold atoms in
optical lattices.  Earlier theoretical work on these systems focused
on novel phenomena within the superfluid phase, where coupling
between fermions and the Bogolubov mode of the bosonic superfluid is
analogous to the electron-phonon coupling in solid state systems.
Several interesting phenomena have been predicted, including
fermionic pairing \cite{albus,LDA,Wang}, charge density
wave order \cite{buechler,Hofstetter,adhikari}, and formation of bound
Fermion-Boson molecules \cite{PowellSachdev,LewensteinSantos}. Yet when Bose-Fermi
mixtures were realized in
experiments\cite{Esslinger,Bongs,BECexp,Bongs2,inguscio,Inguscio-exp,HadziFB}, the most
apparent experimental feature was the dramatic loss of bosonic
coherence in the time of flight experiments even for modest
densities of fermions. This suggested an interesting possibility
that adding fermions can stabilize the Mott states of bosons in
optical lattices. Theoretical work addressing these
experiments, however, suggested that in the case of a homogeneous Bose-Fermi
mixture at constant and low temperatures, the dominant effect of
fermions should be screening of the boson-boson interaction, which
favors the superfluid state \cite{hydro,albus,Kollath}. Hence, the loss of
coherence observed in experiments was attributed to effects of
density redistribution in the parabolic trap or reduced cooling of
the bosons when fermions were added into the mixture.

In this paper we argue that adding fermions into a bosonic system
can actually stabilize bosonic Mott states even for homogeneous
systems. While all previous theoretical analysis represented the
effect of fermions on bosons as an instantaneous screening, in this
paper we take into account retardation effects, which arise from the
presence of very low energy excitations in a Fermi sea. We show that
such retardation gives rise to an effect which is analogous to the
so-called orthogonality catastrophe, which is a
well known cause for X-ray edge singularities and emission
suppression \cite{Wen, Mahan} in solid state systems.

Our paper provides an alternative theoretical
approach to the analysis of the Bose-Fermi mixtures. Rather than doing
perturbation theory from the superfluid state, we consider the Mott
insulating state of bosons as our starting point. This is a
convenient point for developing a perturbation theory, since deep in
the Mott state the bosonic density is uniform and rigid and is
accompanied by a simple Fermi sea of fermions. In general, the SI
transition can be understood as Bose condensation of particle and
hole-like excitations \cite{FWGF} on top of a Mott state. In the absence
of fermions, this condensation requires that the energy cost of
creating particle and hole like excitations, i.e., the Hubbard U, is
compensated by the kinetic energy of these excitations, which is
proportional to both the filling factor and the single particle
tunneling. Adding fermions to the system reduces the energy cost of
creating either a particle or a hole excitation due to
screening \cite{albus,hydro}, but it also reduces the effective tunneling of
bosons. The latter effect can be understood from the following simple
argument. Consider a particle (or a hole) excitation on top of a Mott
state of bosons. For fermions, this extra particle appears as an
impurity on top of a uniform potential. When the bosonic particle
moves to the neighboring site, the
``impurity potential'' for all fermions changes. For individual fermionic
states, the change of the single particle wave function may be
small. But the effective tunneling of the bosonic particle is
proportional to the change of the {\it entire} many-body fermionic
wave function, and therefore we will need to take a product of all single
particle factors. Even when each of the factors is close to one, the product
of many can be much smaller than one. This is the celebrated
``orthogonality catastrophe'' argument of Anderson\cite{AndersonOC}. It can also be
thought of as a polaronic effect in which tunneling of bosons is
strongly reduced due to ``dressing'' by the fermionic screening
cloud. We see that both the interaction and the tunneling are reduced
by adding fermions. It is then a very non-trivial question to
determine which effect dominates, and whether it is the superfluid or
the insulating state that is favored by adding fermions into the
system. Indeed, the main focus of our work is to understand how the
Fermi-Bose system pits the Bosonic superfluidity against the trademark
dynamical effect of free Fermions. A related work, which addresses Fermionic
dynamical effects on the nature of the superfluid-insulator transition, is Ref. \onlinecite{yang-2007}.

In this letter we derive the SF-Mott insulator critical line by
constructing a mean-field theory that contains both the static
screening and dynamical orthogonality catastrophe of the Fermions.
For this purpose we resort to a new path-integral formulation of the
mean-field Weiss theory for the SF-insulator transition \cite{FWGF}.
After demonstrating our approach by deriving the mean-field
transition line for a pure bosonic system, we derive the
path-integral approach to the Fermi-Bose system, and analyze the
results in various limits.

Our analysis will rely on several simplifying assumptions. We
consider only homogeneous systems, which is not the case for
realistic systems in parabolic confining potentials. We do not allow
formation of bound states between particles, which limits us to
small values of the Bose-Fermi interaction strength. The latter
assumption becomes particularly restricting in one dimensional
systems \cite{adhikari,Fleischhauer,MatheyWang,sengupta-2005}, where even small interactions
are effective in creating bound states. We do not take into account
effects of non-equilibrium dynamics, which are important for
understanding behavior of real experimental systems whose parameters
are being changed. And finally, we assume that there are only two
fundamental states for bosons in the presence of fermions:
the superfluid, and the Mott insulator. When our analysis shows proliferation
of particle and hole like excitations inside a Mott state, we
interpret this as the appearance of the superfluid state. We do not
consider the possibility of exotic new phases such as the
compressible state suggested recently by Mering and
Fleischhauer\cite{Fleischhauer}. While these limitations make it
difficult to make direct comparison of our findings to the results
of recent experiments \cite{Esslinger,Bongs,BECexp,Bongs2,inguscio},
we believe that our work provides a new conceptual framework which
can be used to address real experimental systems.

\subsection{Microscopic Model}

The Hamiltonian for the Bose-Fermi system we analyze is given by
\be
\ba{c}
\H=\H_B+\H_F+\H_{int}\vspace{2mm}\\
\H_B=\summ_i[\frac{1}{2}U_B \hat{n}_i^2-\mu
\hat{n}_i-\frac{1}{2}\summ_{\langle i j \rangle}
J \cos\l(\phi_i-\phi_{j}\r)] \vspace{2mm} \\
\H_F=-\summ_{\langle i j\rangle} J_F \cd_i \c_j-\summ_i \mu_F \cd_i
\c_i, \hspace{3mm} \H_{int}=\summ_i U_{FB}\hat{n}_i \cd_i\c_i.
\ea
\label{Heq} 
\ee 
$\H_B$ describes the bosonic gas using the phase and
number operators in each well: $\hat{n}_i,\,\phi_i$. $J$ is the
strength of the Josephson nearest-neighbor coupling (note that $J
\approx n t$, where $n$ is the filling factor and $t$ is the hopping
amplitude for individual bosons), $U_B$ and $\mu$ are the charging
energy and chemical potential, respectively. $\H_F$ describes the
Fermions, with hopping $J_F$ and chemical potential $\mu_F$. The two
gasses have the on-site interaction $U_{FB}$. For simplicity we use
the rotor representation of the Bose-Hubbard model, but our results
are easily generalized to its low-filling limit. The pure Bose gas
forms a superfluid when $J/\Delta\sim 1$, where $\Delta\sim U$ is
the charging gap \cite{FWGF,SachdevBook}. The fermions encourage
superfluidity, on the one hand, by partially screening charging
interactions and reducing the local charge gap \cite{hydro,Kollath}.
But at the same time the fermions' rearrangement motion in response
to boson hopping is slow and costly in terms of the action it
requires. This motion results in an orthogonality catastrophe that
suppresses superfluidity.

Our derivation of the phase diagram is based on the mean-field
approach, which in the case of  purely bosonic systems is equivalent
to the analysis in Refs. \onlinecite{SachdevBook, FWGF, Altman}, but which
can be generalized to study Bose-Fermi mixtures.  The idea is to use
the Weiss approach of reducing the many-site problem in the
Hamiltonian (\ref{Heq}) to a single site problem by assuming the
existence of the expectation value for the phase coherence of
bosons: 
\be 
\langle e^{i\phi_i}\rangle=r. 
\label{mft} 
\ee 
In the local problem, one can calculate a self-consistent equation for $r$
that will produce the transition point. This procedure can not be
simply followed once the fermions are thrown into the mix, since
even with Eq. (\ref{mft}), the Hamiltonian $\H$ is non-local; this
problem is addressed by using the imaginary-time path integral
formulation.

\subsection{Overview}

This paper is organized as follows. In Sec. \ref{nonint} we derive the
path-integral formulation for the mean-field phase boundary of a pure
Bose gas as a function of the parameters in its Hamiltonian and temperature. In Sec. \ref{FB} we build on this formalism to account for
the weakly interacting Bose-Fermi mixture. We find a new condition for
the superfluid insulator transition in terms of the Boson parameters,
as well as the interaction strength, $U_{FB}$, and the Fermion's
density of states, $\rho$. Our main result is
presented in Sec. \ref{resultSec} in Eq. (\ref{rf}). The mean-field
condition is plotted for the cases of fast and slow Fermions, for zero
temperature, as well as at a finite temperature. We conclude the paper with a summary and discussion in Sec. \ref{dis}.

Our main findings are that even a moderately weak interaction with slow Fermions
inhibits superfluidity in the Bosons. The dynamical response of the
slow Fermions produces a large cost in terms of the action for bosonic
number fluctuations. This effect of the orthogonality catastrophe of
a Fermionic screening gas is most apparent where the on-site charging
gap of the Bosons is small ($\Delta\ll U$). In Sec. \ref{resultSec} we also
derive approximate simple expression for the phase boundaries for this
case at zero and low temperatures, Eqs. (\ref{td}) and (\ref{nd}). Our analysis shows that the phase
boundary becomes non-analytical, and superfluidity is dramatically
suppressed.

\section{Pure Bose-gas phase-diagram using the path-integral approach \label{nonint}}

We begin our analysis with the pure bosonic gas. We will use this
case, where no Fermions are present, to derive and demonstrate our
path-integral approach to the mean-field superfluid-insulator
transition. We will first use the mean-field ansatz, Eq. (\ref{mft}), to reduce
the partition function to a path-integral over a single-site
action. Analyzing the single-site action will reveal the mean-field
condition for superfluidity.

The first step is to transform the Hamiltonian $\H_B$
of Eq. (\ref{Heq}) into a single-site Hamiltonian. Using
Eq. (\ref{mft}), we can write:
\be
\H_B\rightarrow \H_{B\,j}=\frac{1}{2}U_B \hat{n}_j^2-\mu
\hat{n}_j-z
\frac{1}{2}J \l(r^{*} e^{i\phi_j}+r e^{-i\phi_j}\r)
\ee
where $z$ is the coordination of the lattice.
The action for site $j$ therefore becomes:
\be
S_j=\int d\tau\l[i\dot{\phi}_j n_j -\frac{1}{2}zJ\l(e^{i\phi_j} r^*+e^{-i\phi_j} r\r)+\frac{1}{2}U {n}_j^2-\mu
{n}_j\r]
\label{mf2}
\ee
Thus the partition function for a single site is:
\be
Z=\int D[\phi(\tau)]\summ_{\{n(\tau)\}}e^{-\intt_0^{\beta}
  d\tau\l[i\dot{\phi} n-z J r\cos\phi +\frac{1}{2}U {n}^2-\mu
{n}\r]}
\ee
where we assumed that $r$ is real, and dropped the index j.

The self-consistent condition for superfluidity equates the degree of
phase ordering on site $j$ with $r$, which was substituted for the
neighbors of site $j$. This mean field equation, Eq. (\ref{mft}), becomes:
\be
\ba{c}
r=\frac{1}{Z}\int D[\phi_j(\tau)]\summ_{\{n_j(\tau)\}}\cos{\phi_j(0)}\exp(-S_j)\vspace{2mm}\\
\approx r\frac{zJ}{Z}\int
D[\phi(\tau)]\summ_{\{n(\tau)\}}\intt_0^{\beta} d\tau_1
\cos{\phi(0)}\cos(\phi(\tau_1)) \vspace{2mm}\\
\exp\l[-\intt_0^{\beta}
  d\tau\l[-in \dot{\phi} +\frac{1}{2}U n^2-\mu n\r]\r].
\label{mf4}
\ea
\ee
where $\cos\phi(0)$ was expanded in $r$ to its lowest power.

Our goal is to simplify condition (\ref{mf4}); for this purpose, we
integrate over the phase variable $\phi$. Let us concentrate first on
the partition function, $Z$, in the denominator of Eq. (\ref{mf4}). In the limit
of $r\rightarrow 0$, $\phi$ only appears in the Berry-phase term,
which using integration by parts becomes:
\be
i\intt_0^{\beta} d\tau n\dot{\phi}=i
n(0)(\phi(\beta)-\phi(0))-i\intt_0^{\beta} d\tau \phi\dot{n}.
\ee
Because $n(0)$ is an integer and $\phi(\tau)$ is periodic on the segment
$[0,\,\beta]$, the first term is always a multiple of
$2\pi i$, and can be omitted. Furthermore, without an $r$ term, the
angle variables in each time slice, $\phi(\tau)$, become Lagrange
multipliers which enforce number conservation in the site:
\be
\int D[\phi(\tau)]\exp(-i\intt_0^{\beta} d\tau
  \phi\dot{n})=\prodd_{\tau} 2\pi
  \delta_{\dot{n}(\tau),\,0}=\prodd_{\tau} 2\pi
  \delta_{n(\tau),\,n(0)}.
\label{delta1}
\ee
Thus $\dot{n}(\tau)=0$, and we can write the single-site partition function as:
\be
Z=\sum_n e^{-\beta\l(\frac{1}{2}U {n}^2-\mu
{n}\r)}.
\ee

Eq. (\ref{mf4})'s numerator is more involved. The phase $\phi(\tau)$
now also appears through the $\cos(\phi(\tau_1))\cos(\phi(0))$
term. A $e^{i\phi(\tau_1)}$ term is indeed a creation operator,
therefore we expect that the cosine factors in the path-integral will change the
number of particles $n(\tau)$ at $\tau_1$ and at $\tau=0$. Let us
demonstrate this by concentrating on the term $\cos\phi(0)=\frac{1}{2}(e^{i\phi(0)}+e^{-i\phi_0})$ and
integrating over $\phi(0)$:
\be
\ba{c}
\int
d\phi(0)\frac{1}{2}\l(e^{-i\phi(0)\dot{n}d\tau+i\phi(0)}+e^{-i\phi(\tau)\dot{n}d\tau-i\phi(0)}\r)=\\
\pi\l(\delta_{\Delta n(0),1}+\delta_{\Delta n(0),-1}\r)
\ea
\ee
where $\Delta n(\tau)=\dot{n}(\tau) d\tau=\mbox{lim}_{\epsilon\rightarrow
  0}n(\tau+\epsilon)-n(\tau-\epsilon)$. The same expression results
from the integration on the $\tau_1$ time slice. Thus the integration of
the $\phi(\tau)$ variables in the numerator of Eq. (\ref{mf4}) still
gives $\dot{n}(\tau)=0$ as long as $\tau\neq 0,\tau_1$. The numerator of
Eq. (\ref{mf4}) can now also be reduced to a simple
sum over $n(\tau)$, but with a jump in the boson-number at $\tau_0$
and $\tau_1$:
\be
n(\tau)=n\pm\theta(\tau)\theta(\tau_1-\tau).
\label{delta2}
\ee

Now that the integration over the $\phi(\tau)$ variables is complete,
we can write the mean-field condition for superfluidity as a single
sum. Since the + and - choices in Eq. (\ref{delta2}) give rise to the
same contribution in the mean-field condition, we can choose the plus,
$n(\tau)=n+\theta(\tau)\theta(\tau_1-\tau)$, and write:
\be
\ba{c}
1=\frac{zJ}{Z}\frac{1}{2}\summ_{n=-\infty}^{\infty}\intt_0^{\beta} d\tau_1 e^{-S[n(\tau)]}=\frac{zJ}{Z}
\summ_{n=-\infty}^{\infty}\intt_0^{\beta} d\tau_1 \\
\frac{1}{2}\exp\l[-\intt_{\tau_1}^{\beta}d\tau \H_{c}(n)\r]\exp[-\intt_0^{\tau_1}d\tau \H_{c}(n+1)],
\label{mf7}
\ea
\ee
with $\H_{c}(n)=\frac{1}{2}Un^2-\mu n$. For a pure Bose gas, we obtain the well
known Weiss mean-field rule for X-Y magnets:
\be
1=\frac{zJ/2}{\summ_{n=-\infty}^{\infty} e^{-\beta \H_c(n)}}\summ_{n=-\infty}^{\infty}\frac{e^{-\beta\H_{c}(n)}-e^{-\beta\H_{c}(n+1)}}{\H_{c}(n+1)-\H_c(n)}.
\label{mf8}
\ee

\section{Effective bosonic action for the SF-insulator transition of the Fermi-Bose mixture\label{FB}}

The addition of Fermions to the Bosonic gas affects the bosons in two
distinct ways. The first is static: the Fermions shift the chemical
potential and the interaction parameters of the Bosons \cite{hydro,albus}. But in the
superfluid phase, Boson number fluctuations become dominant, and the
screening problem becomes a dynamical one. The Fermi screening cloud
requires a finite time to form, and, in addition, it costs an
prohibitively large action in some cases. While the former static screening
effect enhances superfluidity, the latter dynamical effect suppresses
it. The advantage of the imaginary-time path integral formalism, which
was developed in the previous section, is that it deals with both effects
on the same footing, and allows the inclusion of the fermionic
collective dynamical response in a one-site bosonic action.

\subsection{Static screening effects}

Let us now consider the Fermi-Bose mixture of Eq. (\ref{Heq}).
The most straightforward effect of the Fermions is to shift the chemical potential and interaction
parameters. We will first calculate this effect using
a hydrodynamic approach \cite{hydro,albus}. By denoting the DOS of the fermions at
the Fermi-surface as $\rho$, and neglecting its derivative, we can write a charging-energy equation
for the mixture per site:
\be
E_{c}=E_k^{(F)}(n_F)-\mu_F n_F+\frac{1}{2}U n_B^2-\mu_B n_B+U_{FB}n_B n_F
\ee
with
\[
\ba{cc}
\frac{dE_k^{(F)}}{dn_F}=E_F=\mu+\rho^{-1} n_F,&
\frac{d^2E_k^{(F)}}{dn_F^2}=\rho^{-1}.
\ea
\]
By finding the minimum with
respect to the Fermion density, $n_F$, we find:
\be
n_F=n_F^{0}-U_{FB}\rho n_B,
\label{fscreen}
\ee
and the total charging energy is:
\be
E_c=E_0+\frac{1}{2}\l(U-U_{FB}^2\rho\r)n_B^2-\l(\mu_B-U_{FB}n_F^{0}\r)n_B.
\ee
Therefore the charging parameters of the Bose gas are renormalized by
the presence of the Fermions to:
\be
\ba{cc}
\tilde{U}=U-U_{FB}^2\rho & \,\,\tilde\mu_B=\mu_B-U_{FB}n_F^{0}.
\ea
\label{screen}
\ee
This charging energy renormalization makes the Mott lobes
shrink in the $\mu$-$J$ parameter space by the ratio
$U/\tilde{U}$: adding the Fermions mitigates any static charging
effects, since the mobile fermions can screen any local charge even
when the bosons are localized. 

An important note is that in order for the hydrodynamic approach to be
correct, the electronic screening should not exceed one particle. This
restricts the perturbative regime to:
\be
U_{FB}\rho<1.
\label{flimit}
\ee
In addition, for the Fermi-Bose mixture to be stable, we must have
$\tilde{U}>0$, and thus also:
\be
U_{FB}^2\rho<U.
\ee
which in the regime of interest is a less restrictive condition than
Eq. (\ref{flimit}). Next, we consider the Fermionic
dynamical response.

\subsection{The Fermion's dynamical response}

The superfluid bays between the Mott lobes in the traditional
$\mu-J$ phase diagram are affected strongly by a
more subtle and intriguing effect: dynamical screening motion of the
Fermions. The analysis of this effect makes the path integral
necessary. We construct the path integral starting with the action:
\begin{eqnarray} 
\ba{c}
S_{FB}=\int d\tau \l(-i\summ_i (\cd_i
\dot{\c}_i+\n{}_i \dot{\phi}_i) +\H_F+\H_B+\r.\\\l. 
\summ_i
U_{FB}n_F^{0}\n{}_i+U_{FB}\n{}_i\l(\cd_i\c_i-n_F^{0}\r)\r) 
\ea
\end{eqnarray}
with $n_F^{0}=\langle\cd_i\c_i\rangle$, and where $\c$ and $\cd$
should be construed as Grassman variables. The first term in the
second line produces the shift in the chemical potential, as in Eq.
(\ref{screen}), but $U$ is not yet renormalized. The $U$
renormalization is a second order effect, which we analyze by
producing a perturbation series in $U_{FB}\Delta
  n_{F\,i}$, where $ \Delta
  n_{F\,i}=\cd_i\c_i-n_F^0$. An effective action for the Bosons is
  then obtained by integrating over the Fermionic variables:
\be
\ba{c}
e^{-S^{eff}_B}=e^{\int d\tau (i\summ_j \n{}_j\dot{\phi}_j-\tilde{\H}_B)}\cdot\vspace{2mm}\\
\int D[\c]D[\cd]e^{-\int d\tau \l(-i\summ_j\l(\cd_j
  \dot{\c}_j\r)+\H_F+U_{FB}\n{}_j\Delta n_{F\,j}\r)}\approx\vspace{2mm}\\
Z_F e^{\int d\tau (i\summ_j \n{}_j\dot{\phi}_j-\tilde{\H}_B)}\cdot\vspace{2mm}\\
 e^{\frac{1}{2}U_{FB}^2 \int d\tau_1\int d\tau_2
  n_j(\tau_1)\langle \Delta n_{F\,j}(\tau_1) \Delta
  n_{F\,j}(\tau_2)\rangle n_j(\tau_2)},
\ea
\label{newS}
\ee
where $\tilde{\H}_B$ is the pure bosonic Hamiltonian with the
renormalized charging energy and chemical potential, Eq. (\ref{screen}).

From Eq. (\ref{newS}) we see that the integral over the Fermionic
degrees of freedom gives rise to a new Boson interaction term. It is given as a
polarizability bubble for the fermions:
\be
\ba{c}
\langle \Delta n_{F\,i} \Delta n_{F\,i}(0)\rangle_{\omega}=T \frac{1}{V^2}\summ_{\vec{k}_1,\,\vec{k}_2}\vspace{2mm}\\
\summ_{\omega'}\frac{-1}{(i\omega'-\xi_{\vec{k}_1})(i(\omega'+\omega)-\xi_{\vec{k}_2})},
\ea
\label{prepol}
\ee
with $\xi_{\vec{k}}=\epsilon_{\vec{k}}-\mu_F$ being the fermionic
kinetic energy relative to the Fermi surface. After
some manipulations (see App. \ref{appA}) we obtain:
\be
\langle \Delta n_{F\,i} \Delta
  n_{F\,i}(0)\rangle_{\omega}\approx \l\{\ba{cr}
a\rho-\pi|\omega|\rho^2 & |\omega|<\rho^-1 \\ c/\omega^2 &
|\omega|>1/\rho,
\ea\r.
\label{pol}
\ee
and as we discuss below, the perturbative analysis is
  valid when $\rho U_{FB}<1$. The first low-frequency term in Eq. (\ref{pol}) yields the static screening, Eq. (\ref{screen}), i.e.,
$a=U_{FB}^2\rho$. The second term describes the dynamical response, producing the action term:
\be
S_{OC}=T\summ_{\omega}\dot{n}_{\omega}\dot{n}_{\omega}^*\frac{\pi U_{FB}^2\rho^2}{2|\omega|}
\label{OC}
\ee
This term yields logarithmic contributions to the action, whose
effects are familiar from electronic systems as orthogonality catastrophe in metal X-ray
absorption spectrum,\cite{AndersonOC,Nozieres} the Kondo effect \cite{AndersonYuval}, and
Caldeira-Leggett dissipation \cite{C-L1,C-L2}. Here its effect is to
suppress superfluidity, since it couples to the number
fluctuations. At angular frequencies greater than $1/\rho$, the
interaction term decays quickly ($c$ is a positive constant), and hence $\Lambda=1/\rho$ serves as
a UV cutoff. It also implies that the screening and logarithmic
contributions to the action can only appear with a time-lag $\tau_d\sim\rho$.

The next step is to calculate the action $S[(n(\tau)]$ with $n(\tau)=n+\theta(\tau)\theta(\tau_1-\tau)$ as in
Eq. (\ref{mf7}). We have:
\be
S[n(\tau)]=S_{charging}+S^{\Lambda}+S^{OC}
\ee
First, $S_{charging}=\int_0^{\beta}(\frac{1}{2}\tilde{U}
n(\tau)^2-\tilde{\mu}n(\tau))$.  But the
cutoff $\Lambda\approx 1/\rho$ of the fermions' polarizability
implies that the screening cloud forms only after the time $\rho$. The
instantaneous screening is thus modified by the action (when $T\rho\ll
1$):
\be
\ba{c}
S^{\Lambda}(\tau_1)\approx\\
\frac{U_{FB}^2\rho}{2}(\tau_1-\rho\ln[\cosh\frac{\tau_1}{\rho}\frac{\cosh\frac{\beta-\tau_1}{\rho}\cosh\frac{\beta+\tau_1}{\rho}}{\cosh^2{\frac{\beta}{\rho}}}]).
\label{retard}
\ea\ee
While $S_{charging}$ assumes that the polarizability, Eq. (\ref{pol})
has the term $a\rho$ for all frequencies, the correction term
$S^{\Lambda}$ takes into account the cutoff in the static screening
term. Instead of the fermionic screening in the wake of a change of
$n_B$ at $\tau=0$ being $\Delta n_F\sim \theta(\tau)$, we have
$\Delta n_F\sim \mathrm{arctanh}(\tau/\rho)$. Considering in addition the
periodic nature of imaginary time, we obtain Eq. (\ref{retard}).

Finally, $S_{OC}$, contains the contribution due to the orthogonality catastrophe:
\be
S_{OC}^{0\rightarrow \tau_1}=T\summ_{\omega}[\cos(\omega\tau_1)-1]\frac{\pi U_{FB}^2\rho^2}{|\omega|}=\gamma\ln\frac{\sin
  (\pi T\tau_1)}{\sin
  (\pi T/\Lambda)}.
\label{OC1}
\ee
In Eq. (\ref{OC1}) we defined the dissipation parameter:
\be
\gamma=U_{FB}^2\rho^2.
\label{gamma}
\ee
Note that we are restricted to $\gamma<1$ in the perturbative regime,
Eq. (\ref{flimit}).

\section{Mean-field phase diagram and the orthogonality catastrophe \label{resultSec}}

The mean-field transition line is obtained, as in Sec. \ref{nonint},
from Eq. (\ref{mf7}), which here takes the form:
\be
\ba{c}
1=\frac{zJ}{2Z}\summ_{n=-\infty}^{\infty}\intt_0^{\beta} d\tau_1 e^{-S[n(\tau)]}
\ea
\ee
Substituting $S[n(\tau)]=S_{charging}+S^{\Lambda}+S^{OC}$ from
Eqs. (\ref{retard}) and (\ref{OC1}), we obtain the mean-field
condition for the transition line:
\be
\ba{c}
1=\frac{zJ}{Z}(2\pi)^N
\summ_{n=-\infty}^{\infty}\int_{0}^{\beta} d\tau
\frac{1}{2}\exp\l[-\beta \tilde{\H}_{c}(n)\r]\\
\exp[-\tau
  (\tilde{\H}_{c}(n+1)-\tilde{\H}_{c}(n))-S^{\Lambda}(\tau)]\cdot \l(\frac{\sin
  (\pi T/\Lambda)}{\sin
  (\pi T\tau)}\r)^{\gamma},
\label{rf}
\ea
\ee
with $\tilde{\H}_{c}(n)=\frac{1}{2}\tilde{U}n^2-\tilde{\mu} n$, and
$S^{\Lambda}(\tau)$ given in Eq. (\ref{retard}). This is our main result.

\begin{figure}
\includegraphics[width=7.5cm]{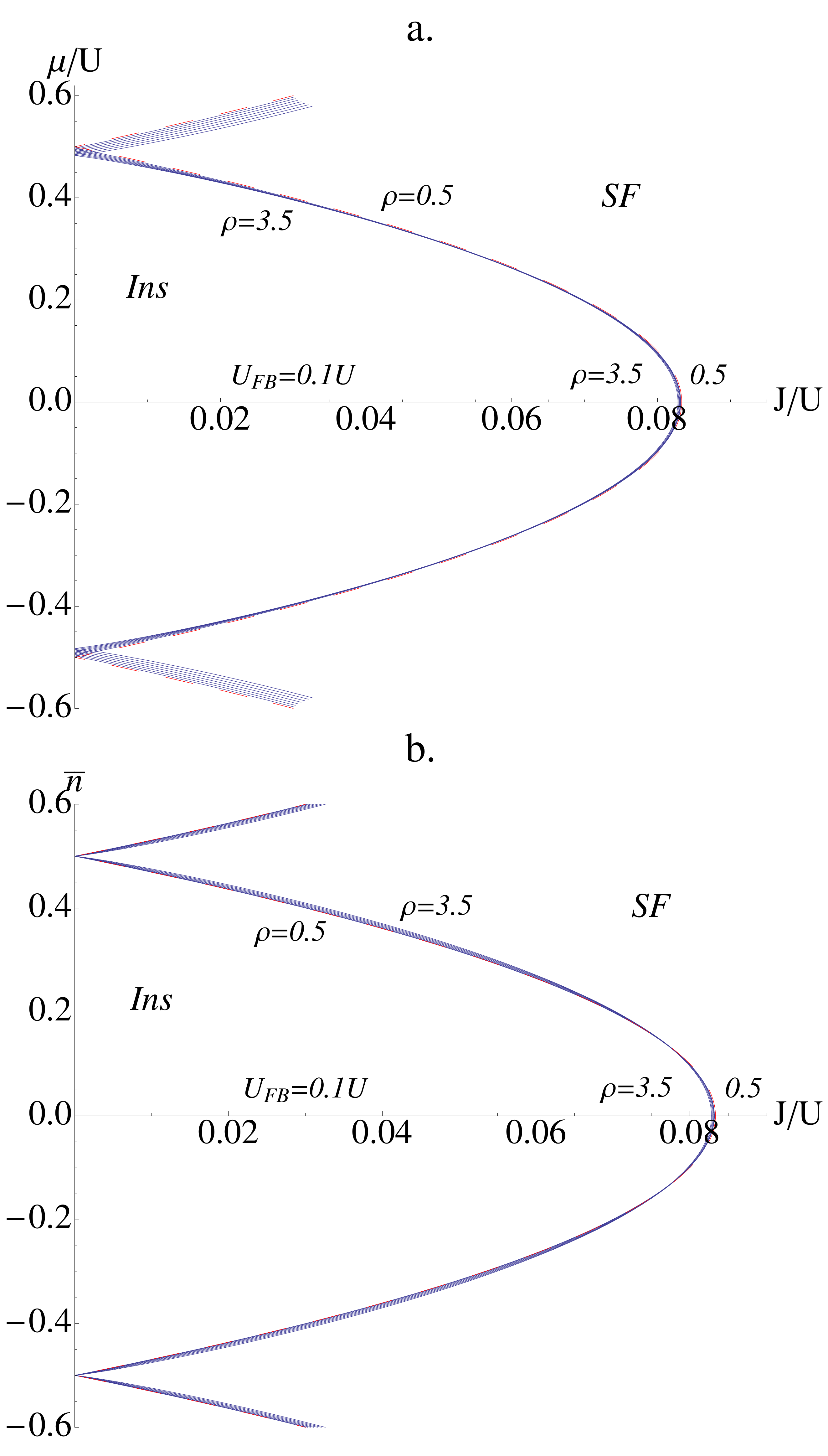}
\caption{Zero-temperature mean-field phase diagram of the Fermi-Bose
  mixture with $U_{FB}=0.1U$. {\bf (a)} Renormalized chemical
  potential, $\tilde{\mu}/U$ vs. bare $J/U$ phase
  boundary. From bottom-left to top-right, the fermions DOS is
  $\rho\cdot U=3.5,3,2.5,\ldots,0.5$. A dashed red line marks the
  uncoupled Bose gas, $U_{FB}=0$, but can barely be distinguished from
  the $\rho=0.5$ line. {\bf (b)}  charge-offset $\nbar=\tilde{\mu}/\tilde{U}$
  vs. bare $J/U$ phase boundary. From left to right at $\nbar=0$, the fermions DOS is
  $\rho\cdot U=3.5,3,2.5,\ldots,0.5$. \label{weakT0}}
\end{figure}

\begin{figure}
\includegraphics[width=7.5cm]{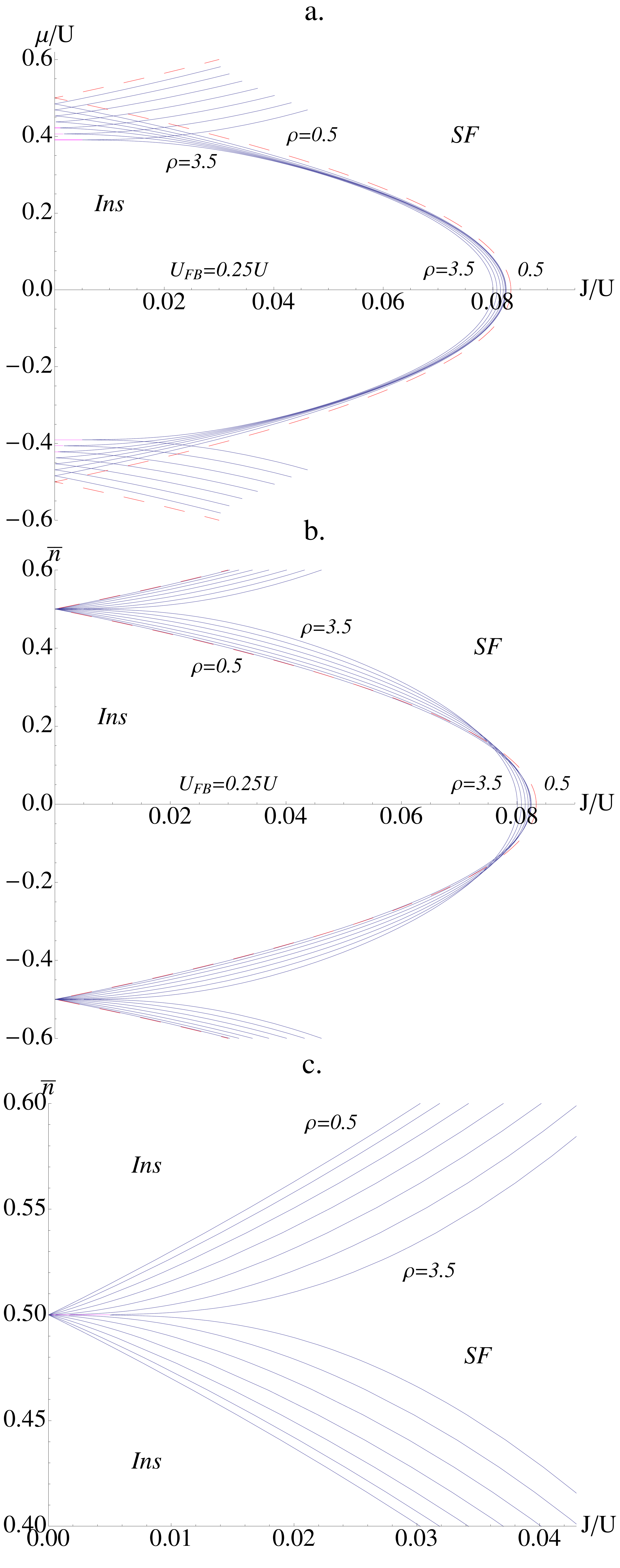}
\caption{Zero-temperature mean-field phase diagram of the Fermi-Bose
  mixture with $U_{FB}=0.25U$. {\bf (a)} Renormalized chemical
  potential, $\tilde{\mu}/U$ vs. bare $J/U$ phase
  boundary. From bottom-left to top-right, the fermions DOS is
  $\rho\cdot U=3.5,3,2.5,\ldots,0.5$. The dashed red line is the
  uncoupled Bose gas, $U_{FB}=0$. {\bf (b)}  charge-offset $\nbar=\tilde{\mu}/\tilde{U}$
  vs. bare $J/U$ phase boundary. From left to right at $\nbar=0$, the fermions DOS is
  $\rho\cdot U=3.5,3,2.5,\ldots,0.5$. {\bf (c)} A focus on the area
  near degeneracy, $\nbar=1/2$, in (b). In this plot we can see the
  strong effect of the orthogonality catastrophe for slow
  electrons. The intercept of the boundary with the $\nbar$ axis
  becomes singular and scales as
  $J/U\sim|0.5-\nbar|^{1-\gamma}$, illustrating Eq. (\ref{nd}). 
\label{strongT0}} 
\end{figure}

\begin{figure}
\includegraphics[width=7.5cm]{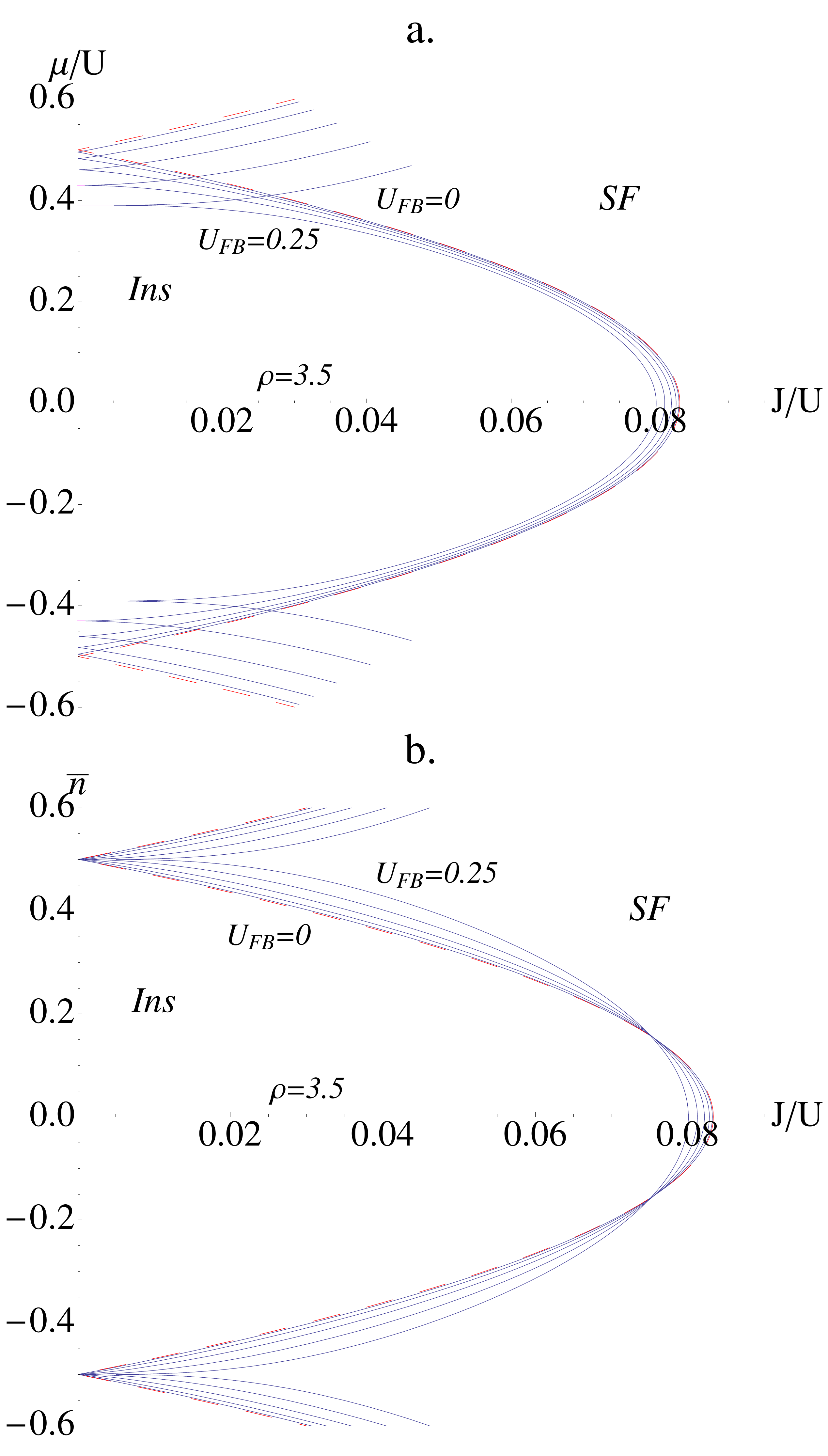}
\caption{Zero-temperature mean-field phase diagram of the Fermi-Bose
  mixture with Fermi dispersion $\rho=3.5/U$. {\bf (a)} Renormalized chemical
  potential, $\tilde{\mu}/U$ vs. bare $J/U$ phase
  boundary. From bottom-left to top-right, the fermions DOS is
  $U_{FB}/U=0,0.05,0.1,\ldots,0.25$. $U_{FB}=0$ is shown as a dashed
  red line. {\bf (b)} Charge-offset $\nbar=\tilde{\mu}/\tilde{U}$
  vs. bare $J/U$ phase boundary. From left to right at $\nbar=0$, the fermions DOS is
  $U_{FB}/U=0.25,0.2,\ldots,0$, with $U_{FB}=0$ shown as a red dashed line. \label{slowT0}}
\end{figure}

\begin{figure}
\includegraphics[width=7.5cm]{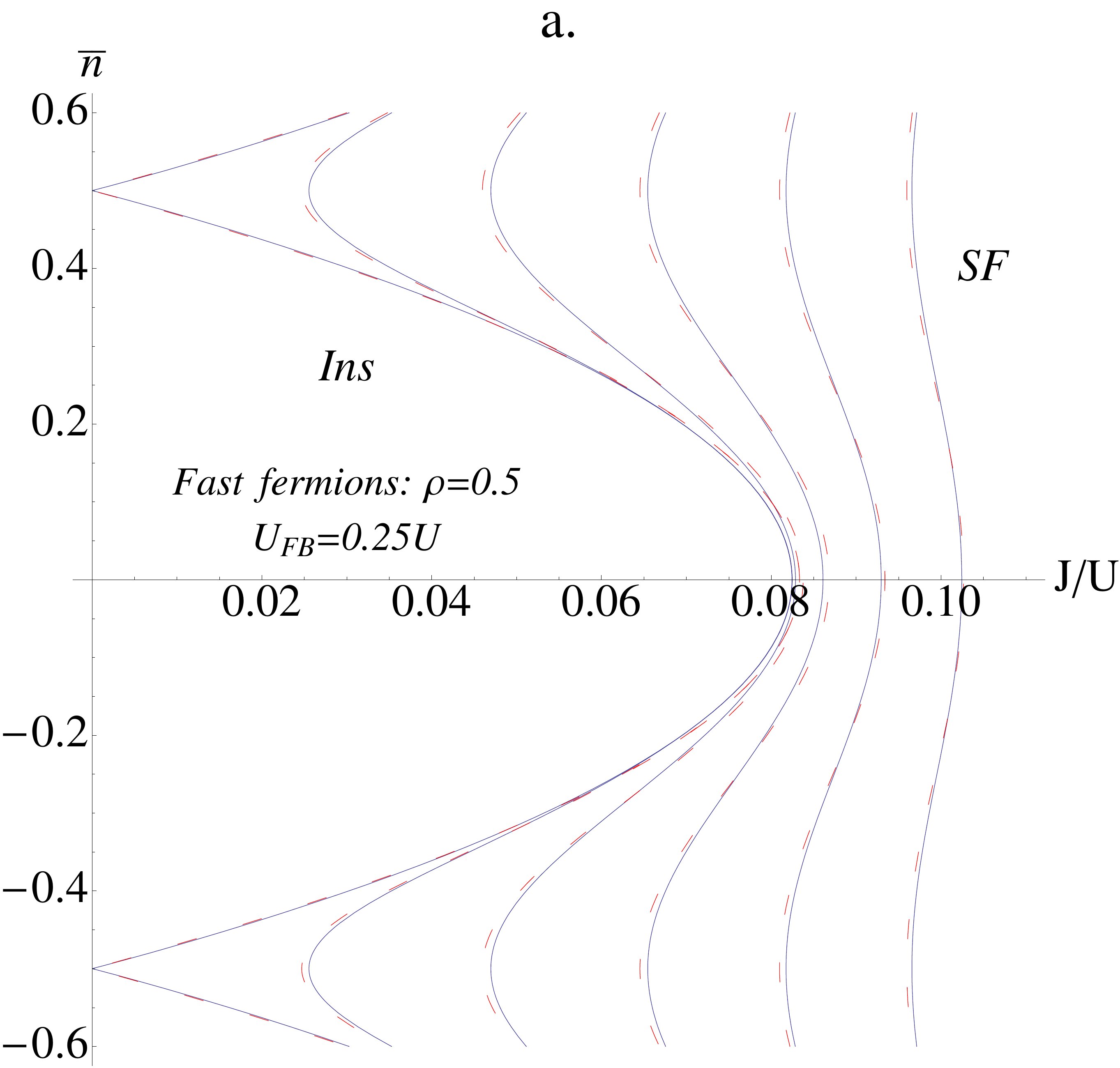}
\includegraphics[width=7.5cm]{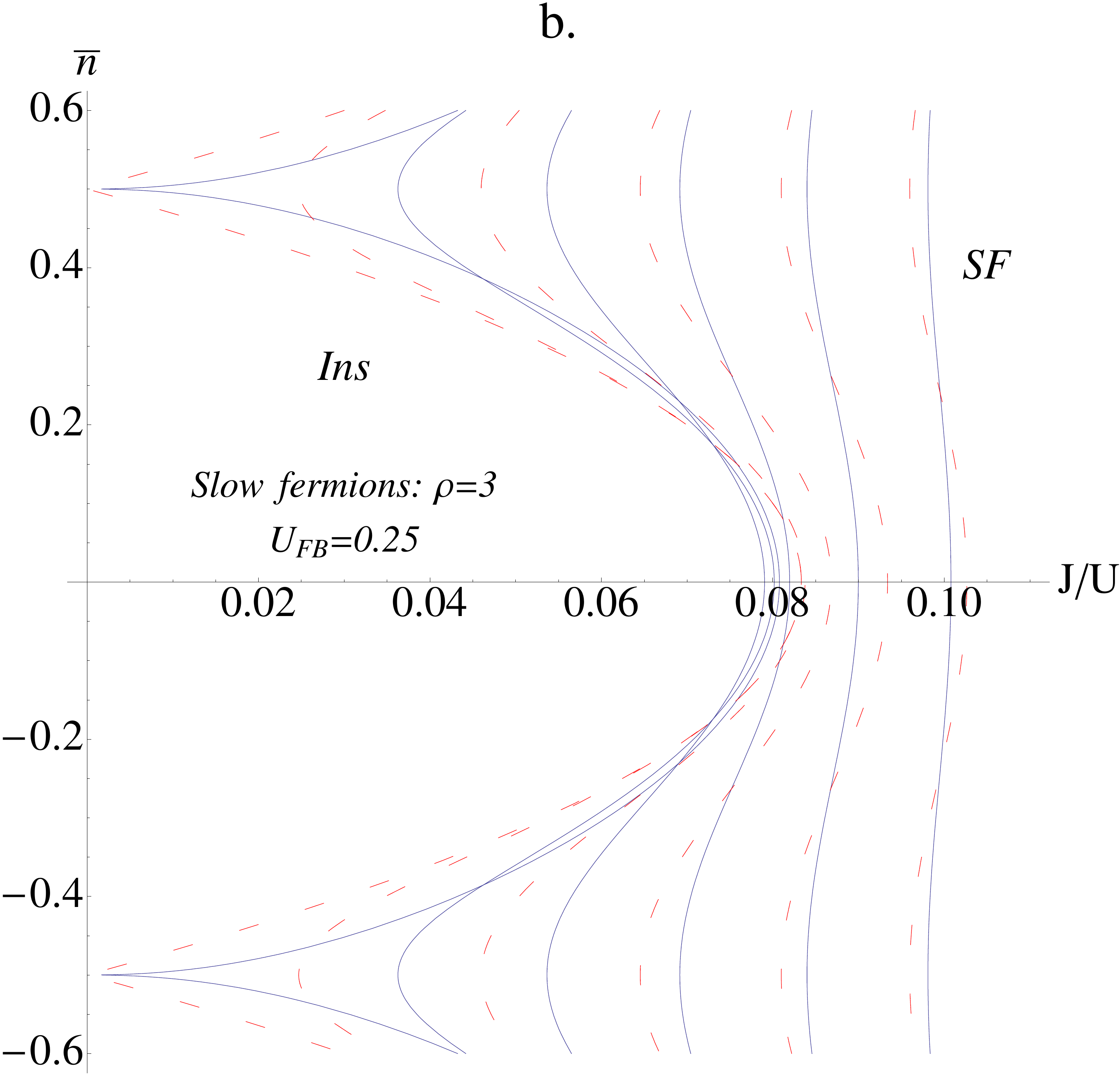}
\includegraphics[width=7.5cm]{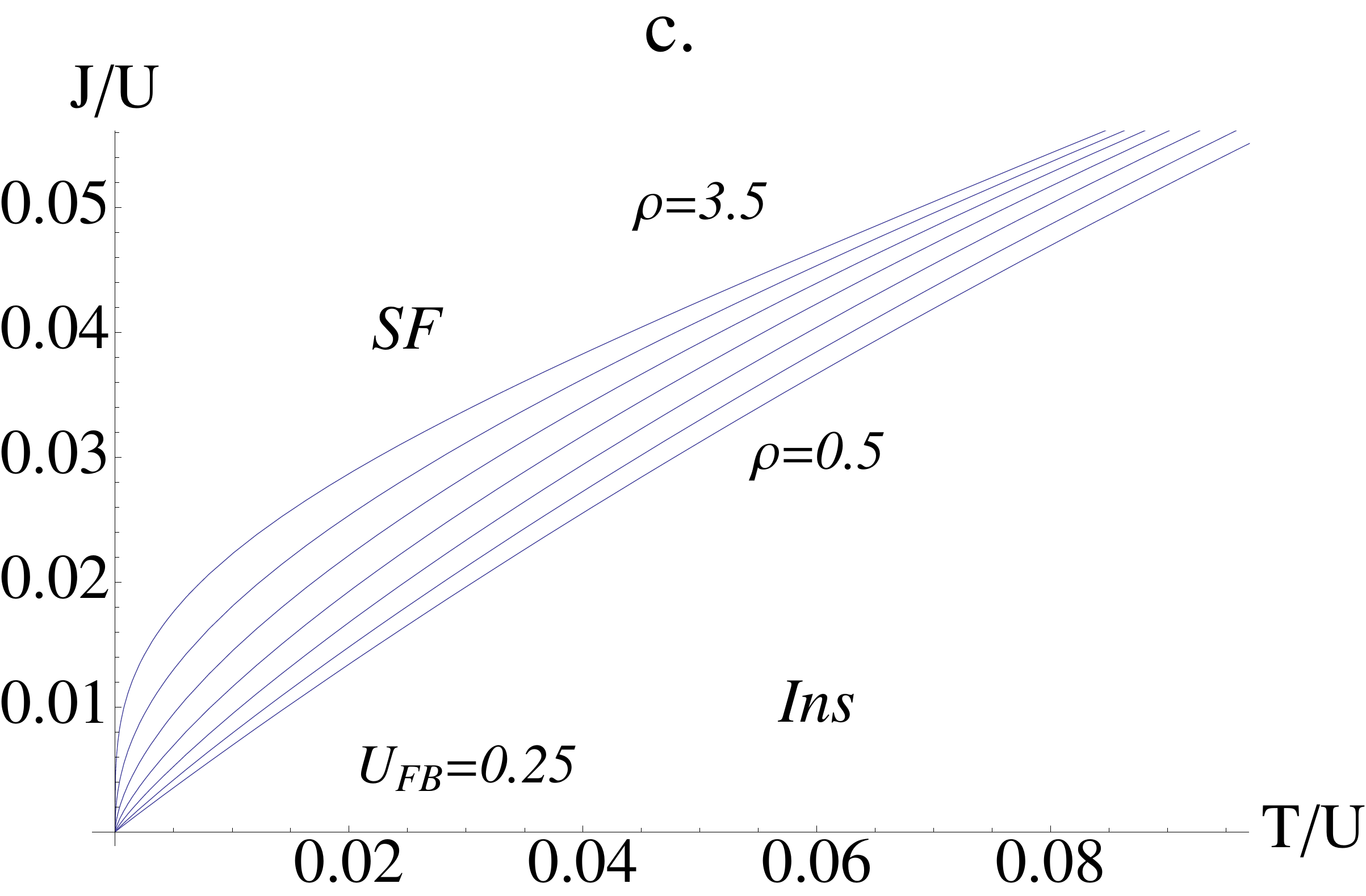}
\caption{Temperature dependence of the SF-Ins boundary for $U_{FB}=0.25U$. {\bf (a)} The
  SF-Ins boundary for $T=0,0.04,\ldots,0.2$ (from left to right at
  $\nbar=1/2$) with
  fast fermions, $\rho=0.5$. The dashed red line is the uncoupled Bose gas
  boundary, $U_{FB}=0$. {\bf (b)} Same as (a) with slow Fermions,
  $\rho=3/U$.  {\bf (c)} The critical boson hopping $J/U$
  vs. $T/U$, for fermion DOS $\rho\cdot
  U=0.5,1,\ldots,3.5$. The linear curve of the uncoupled Bose gas
  becomes a cusp as the bosons couple to slow fermions, as also
  calculated in Eq. (\ref{td}). \label{Tplot}}
\end{figure}

Eq. (\ref{rf}) allows us to calculate the mean-field SF-insulator
phase boundary for weakly interacting mixtures for a range of
Temperatures and Fermi DOS. To illustrate Eq. (\ref{rf}) predictions for the
transition line, Figs. \ref{weakT0}, and \ref{strongT0} show the
boundaries for Bosons and Fermions interacting with $U_{FB}=0.1U$ and
$U_{FB}=0.25U$, respectively, for a range of fermion velocities, or
DOS $\rho$. In Fig. \ref{slowT0} we plot the effect of slow fermions,
with $\rho=3.5/U$, on the Bosonic SF transition for a range of
interactions $U_{FB}$. The Superfluid-insulator transition boundary at
finite temperature for $U_{FB}=0.25U$ is shown in Fig. \ref{Tplot}
for a range of temperatures for fast and slow fermions. In all figures
we assume $z=6$.

One easily drawn qualitative conclusion is that slow electrons mostly inhibit superfluidity, which is the
mark of the orthogonality catastrophe. The  most dramatic suppression effect occurs near the degeneracy
points, where $\nbar=\tilde{\mu}/\tilde{U}=m+1/2$. Let us obtain closed-form expressions for the SF-INS boundary
there.  A helpful observation is that if the charging gap nearly vanishes, it suffices to
consider the lowest nearly-degenerate charge states in Eq. (\ref{rf}).

When the degeneracy is exact, e.g., at $\nbar=1/2$, we
obtain for the critical $J$ vs. Temperature:
\be
\ba{c}
1=\frac{zJ}{4}\intt_{0}^{\beta}d\tau\l(\frac{\sin(\pi
  T/\Lambda)}{\sin(\pi T \tau)}\r)^{\gamma}\approx\\
 \frac{zJ}{4\pi
  T}\l(\frac{2\pi
  T}{\Lambda}\r)^{\gamma}B(\frac{\gamma}{2},1-\gamma)\sin\l(\frac{\pi\gamma}{2}\r)
\label{td}
\ea
\ee
where we neglected screening retardation, i.e., $S^{\Lambda}$,
altogether. This is valid for $T^{1-\gamma}\ll \Lambda^{1-\gamma}/\gamma$. $B(m,n)$ is the
beta-function. Fig. \ref{Tplot}b shows Eq. (\ref{rf}) in this limit.
Whereas for pure bosons the critical $J$ is linearly proportional to
$T$, the interaction with Fermions makes the critical $J$ required for
superfluidity increase dramatically, and obey $J_c\sim T^{1-\gamma}$.

A similar analysis can be done at zero temperature slightly away from
degeneracy at $\nbar-1/2=\epsilon\ll 1$. In this regime we obtain:
\be
1=\frac{zJ}{4}\intt_{0}^{\infty}d\tau\frac{e^{-\epsilon\tilde{U}\tau}}{\l(\Lambda \tau\r)^{\gamma}}\approx\\
 \frac{zJ}{4}\l(\frac{\epsilon
  \tilde{U}}{\Lambda}\r)^{\gamma}\frac{1}{\epsilon\tilde{U}}\Gamma(1-\gamma).
\label{nd}
\ee
where ignoring screening retardation is valid if
$\epsilon^{1-\gamma}\ll (\Lambda/\tilde{U})^{1-\gamma}/\gamma$. Here
too, the critical hopping as a function of $\epsilon$ is linear,
$J_c\sim |\epsilon|$ for pure Bosons, but increases dramatically to
$J_c\sim |\epsilon|^{1-\gamma}$ when the Bosons interact with the
Fermions. This dependence is demonstrated in Fig. \ref{strongT0}c.

\section{Discussion \label{dis}}

\subsection{Regime of applicability}

Our theory of the SF-insulator transition applies to the weakly coupled Bose-Fermi
mixtures, with $U_{FB}< U$, but a second parameter that is required to
be small is $U_{FB}\rho<1$ (see Eq. \ref{flimit}). The latter is required for the
perturbation theory of Eq. (\ref{newS}) to be justified.
As explained above, this condition can be easily understood by noting from
  Eq. (\ref{fscreen}) that the response of the Fermi gas to the appearance
  of a Boson in a particular site is $\Delta n_F=U_{FB}\rho$, which
  clearly must be lower than 1. Even more importantly, the
  perturbative analysis is valid so long that no localized states
form in the fermionic spectrum when a site's potential
changes by $U_{FB}$; this, too, is true when $\rho
  U_{FB}<1$ at large dimensionality.

The formation of a localized state at larger values of $U_{FB}\rho$,
and therefore where $\gamma>1$, is likely to
suppress the orthogonality-catastrophe effects, perhaps in analogy
to the behavior of a Kondo-impurity in a metal: when a Kondo
impurity localizes an electronic state it becomes inert.
Therefore the largest suppression of the SF-INS boundary is
likely to occur when $U_{FB}\rho\sim 1$, as $\gamma\sim
1$\cite{footnoteX}. The regime $\gamma>1$ lies beyond the scope of this paper,
but we intend to approach it in a later publication. Note that this
regime can still occur when $U_{FB}\ll U$.

We note also that since our theory is concerned with weak Fermi-Bose coupling, it ignores Fermi-Bose bound
pair formation, as well as p-wave
superconducting correlations, which may be important only at
parametrically low temperatures.

\subsection{Relation to experiment}

Our theory provides the mean-field phase diagram under the
assumption of a grand-canonical ensemble with fixed chemical
potentials. Experiments, on the other hand, are conducted in finite
non-uniform traps, and therefore to compare their results with the
thermodynamic phase diagram we provide, the chemical potential of the interacting Bose and Fermi gasses
must be determined for particular trap geometries, using, e.g., the LDA
approach, as in Refs. \onlinecite{LDA,albus}.

Experiments on Bose-Fermi systems show a strong
suppression of superfluidity. Ref. \onlinecite{Esslinger} describes a
system where $J/U\sim 1/20$, $J_F/J\sim
5$, and $U_{FB}\sim -2 U$, i.e., the Bose-Fermi system is strongly
interacting. Thus our theory of orthogonality-catastrophe effects
is not directly applicable here. We note, however,
that at large values of $U_{FB}$, bound composite fermions would
form\cite{LewensteinSantos}. These will have a weakened $U_{FB}$ and a strongly
enhanced DOS, ${\rho}$. This makes it possible to observe
orthogonality-catastrophe SF suppression even in this regime.

\subsection{Relation to dissipative phase transitions}

At weak $U_{FB}$, we find that Fermions, through their orthogonality catastrophe, by and large {\it inhibit}
superfluidity, particularly when the fermions are slow. This effect is
extremely reminiscent of dissipative superconducting-metal phase transitions in
Josephson junctions.

Typically, dissipative
effects as in resistively shunted Josephson
junctions (RSJJ), are thought to strengthen phase coherence
\cite{Schmid, Chakravarty}. But in our case, since the Bose-Fermi mixtures
couple through a capacitive interaction, as opposed to the
phase-phase interaction in superconducting systems expressed in a Caldeira-Leggett
\cite{C-L1,C-L2} term or its modular equivalent \cite{AES}, we
encounter a suppression.

Another important distinction is that
the Caldeira-Leggett analysis of a single RSJJ's, and of 1d
superfluids, associates (Quasi long-range) phase ordering with the
long-time behavior of $\langle e^{i\phi(\tau)}e^{-i\phi(0)}\rangle$.
But in the mean-field theory of the SF-insulator transition, the onset
of {\it true} long-range phase order we encounter is associated with
the less restrictive {\it time-integral} of the aforementioned correlation, as in Eq. (\ref{rf})
\cite{footnote1}. The dynamics of the Fermionic screening gas modifies
this integral only quantitatively, but it does not affect the nature
of the transition. 

The less restrictive condition for ordering in
the mean-field analysis reflects the assumed higher dimensionality of
the systems we consider. Concomitantly, in low dimensional systems
with short-range interactions only, the Mermin-Wagner theorem rules
out the formation of long range order. For Josephson junctions, for instance,
phase-slips are the domain-wall like defects which make the order parameter fluctuate. Similar defects are absent from our
analysis since their cost in terms of action is too prohibitive
due to the assumed high connectivity of the system. Therefore we can
make the mean-field assumption of a non-fluctuating order
parameter. This assumption is fully justified above the lower critical
dimension (although our analysis will only be valid at and above the
upper critical dimension).

\subsection{Summary and future directions}

In this manuscript we concentrated on the effects of the orthogonality
catastrophe on the superfluid-insulator transition line, and showed how slow fermions inhibit
superfluidity through dissipation capacitatively coupled to number-fluctuations. The
orthogonality catastrophe should also be
evident in other measurements, which may give an independent
estimate of the dissipation present. This might be most apparent in
revival experiments, where the system is shifted from a superfluid
phase into the insulating phase, and released after $t_{w}$
\cite{Bloch-revival}. We expect that the revival decay time will
be smaller with increasing dissipation. We intend to address the
dynamical effects in Fermi-Bose mixtures in a future work.

Another interesting angle for future work
is the appearance of a super-solid at the special point of Fermionic
half-filling \cite{Hofstetter}, extending our formalism to account for
this possibility could be done by considering the Fermionic density
correlations near nesting vectors of the Fermi gas.

\acknowledgments

We gratefully acknowledge useful discussions with E. 
Altman, H.P. B\"uchler, I. Bloch, T. Esslinger, W. Hofstetter, M. Inguscio, W. Ketterle, 
and R. Sensarma. This work was supported by AFOSR, DARPA, 
Harvard-MIT CUA, and the NSF grant DMR-0705472. 

\appendix

\begin{widetext}
\section{Fermionic response function \label{appA}}

For completeness, we provide here a simple derivation of the
Fermionic response function, Eq. (\ref{pol}), for fermions in
one-dimension. Once our result is put in terms of the fermionic
density-of-states at the Fermi-surface, it applies in any
dimension, since the existence of a $(d-1)$-dimensional Fermi-surface
renders the dispersion for low-energy excitations essentially
one-dimensional. 

Let us assume, for simplicity, that the fermionic Hamiltonian is:
\be
\H=\hbar v_F\summ_{k=0}^{2k_F} (|k|-k_F)(\cd_k\c_k+\cd_{-k}\c_{-k})
\label{fermiH}
\ee
The density of states per site for this Hamiltonian is:
\be
\rho=\frac{1}{\pi\hbar v_F}
\ee
For the Hamiltonian (\ref{fermiH}), Eq. (\ref{prepol}) reads:
\be
C_{\omega}=\langle \Delta n_{F\,i} \Delta n_{F\,i}(0)\rangle_{\omega}=T \frac{1}{V^2}\summ_{k_1,k_2=-2k_F}^{k_F}\summ_{\omega'=2\pi(n+1/2)}\frac{-1}{(i\omega'-\hbar v_F
  (|k_1|-k_F))(i(\omega'+\omega)-\hbar v_F (|k_2|-k_F))},
\ee
This formula sums over the contributions of particle-hole excitations
of four kinds: both particle and hole are right movers ($k_1,k_2>0$),
both particle and hole are left movers ($k_1,k_2<0$), and two mixed
cases. In order to avoid the absolute value, we concentrate on the
first case, and multiply by four:
\be
C_{\omega}=4 T\summ_{\omega'=2\pi(n+1/2)+\omega/2}\int_0^{2k_F}\frac{dk_1}{2\pi}\int_0^{2k_F}\frac{dk_2}{2\pi}\frac{1}{i(\omega'-\omega/2)-\hbar v_F
  (k_1-k_F)}\frac{1}{i(\omega'+\omega/2)-\hbar v_F (k_2-k_F)},
\ee
where we also shifted $\omega'$ by $\omega/2$. We now separate from this sum the contributions from large $|\omega'|>\hbar
v_F k_F$. These high energy modes contribute an $\omega$-independent
term to the static screening. Corrections to this constant are easily
seen to be quadratic in $\omega$ (e.g., set $k_1,k_2\rightarrow
0$). The low-$\omega'$ terms, however, will give rise to a $|\omega|$
contribution, which we are after. The k-integrals can be easily done
in the limit $|\omega'\pm\omega/2|\ll \hbar v_F k_F$, and we obtain:
\be
\ba{c}
C_{\omega}=c-4T\summ_{\omega'=2\pi(n+1/2)+\omega/2} \frac{1}{(2\pi\hbar
  v_F)^2}\log \l(\frac{\hbar v_F k_F-i(\omega'-\omega/2)}{-\hbar v_F
  k_F-i(\omega'-\omega/2)}\r)
\log \l(\frac{\hbar v_F k_F-i(\omega'+\omega/2)}{-\hbar v_F k_F-i(\omega'+\omega/2)}\r)\vspace{2mm}\\
\approx c-4T\summ_{\omega'=2\pi(n+1/2)+\omega/2}\frac{1}{(2\pi\hbar
  v_F)^2} [-i\pi~\rm{sgn}(\omega'+\omega/2)][-i\pi~\rm{sgn}(\omega'-\omega/2)].
\ea
\ee
The $\omega$ dependence arises from the region where the two sign
functions give opposite results:
$-|\omega|/2<\omega'<|\omega|/2$. Thus:
\be
C_{\omega}=c'-4T \frac{|\omega|}{2\pi T}\frac{1}{(2\hbar v_F)^2}\cdot
2,
\ee
where the last factor of 2 is since $c'$ contains the contributions
for $-|\omega|/2<\omega'<|\omega|/2$ assuming the same sign as for the
rest of the frequency range. The final answer is thus:
\be
C_{\omega}=c'-\frac{|\omega|}{\pi v_F^2\hbar^2}=a-\pi\rho^2|\omega|,
\ee
as reported in Eq. (\ref{pol}).

\subsection{Bosonization approach to the polarization calculation}

One-dimensional fermionic systems are most effectively described in
terms of a bosonized action. Let us re-derive Eq. (\ref{pol}) using
this simpler approach. We define the two fields $\theta$ and
$\phi$. Using here the convention:
\be
\frac{1}{\pi}\nabla\theta=\rho_L+\rho_R \hspace{5mm} \frac{1}{\pi}\nabla\phi=\rho_R-\rho_L,
\ee
where $\rho_{R,L}$ are the right-moving and left-moving densities
respectively, the Hamiltonian of 1-d Fermions is:
\be
\H=\frac{v_F\hbar}{2\pi}\int dx \l(\nabla\theta^2+\nabla\phi^2\r),
\label{hbose}
\ee
and $v_F$ is the Fermi velocity. 

The Hamiltonian (\ref{hbose}) can be turned into an imaginary time
Lagrangian:
\be
{\cal L}= \frac{v_F\hbar}{2\pi}\int dx
\l(\nabla\theta^2+\dot{\theta}^2\r).
\label{Lbose}
\ee
The density-density correlation we would like to calculate is now
given as a path-integral over the $\theta$ field:
\be
C_{\omega}=\langle \Delta n_{F\,i} \Delta
n_{F\,i}(0)\rangle_{\omega}=2\cdot\frac{1}{\pi^2}\langle\nabla\theta\nabla\theta\rangle_{\omega}
\ee
Here we need to pause and explain the extra factor of 2: the
expectation value in the brackets only takes into account
particle-hole excitations that are contained within the same branch of
the fermionic spectrum,
right moving or left moving. We must also include, however, excitations with
the particle part being a right mover and the hole being a left mover, and
vise versa. These give exactly the same contribution (as is also seen
in the first approach), and therefore it is sufficient to simply
multiply the expectation value by 2. 

With that in mind, we proceed to write:
\be
\ba{c}
C_{\omega}=\frac{2}{\pi^2}\intt_{-k_F}^{k_F}
\frac{dk}{2\pi}\frac{k^2}{\frac{2v_F\hbar}{2\pi}\l(k^2+\omega^2/v_F^2\r)}\vspace{2mm}\\
\frac{1}{\pi^2v_F \hbar}\intt_{-k_F}^{k_F} dk\l[ 1-\frac{\omega^2/v_F^2}{k^2+\omega^2/v_F^2}\r]\vspace{2mm}\\
\approx c' -\frac{\omega^2}{\pi^2v_F^3 \hbar}\intt_{\infty}^{\infty}
\frac{dk}{k^2+(\omega/v_F)^2}\vspace{2mm}\\
=c'-|\omega|\frac{1}{\pi v_F^2\hbar^2}=c'- \pi\rho^2 |\omega|,
\ea\ee
where in the last step we simply calculated the residue of the k
integral. 
\newline
\newline

\end{widetext}

\bibliography{FB}

\begin{thebibliography}{58}
\expandafter\ifx\csname natexlab\endcsname\relax\def\natexlab#1{#1}\fi
\expandafter\ifx\csname bibnamefont\endcsname\relax
  \def\bibnamefont#1{#1}\fi
\expandafter\ifx\csname bibfnamefont\endcsname\relax
  \def\bibfnamefont#1{#1}\fi
\expandafter\ifx\csname citenamefont\endcsname\relax
  \def\citenamefont#1{#1}\fi
\expandafter\ifx\csname url\endcsname\relax
  \def\url#1{\texttt{#1}}\fi
\expandafter\ifx\csname urlprefix\endcsname\relax\def\urlprefix{URL }\fi
\providecommand{\bibinfo}[2]{#2}
\providecommand{\eprint}[2][]{\url{#2}}

\bibitem[{\citenamefont{Haviland et~al.}(1989)\citenamefont{Haviland, Liu, and
  Goldman}}]{GoldmanHaviland}
\bibinfo{author}{\bibfnamefont{D.~B.} \bibnamefont{Haviland}},
  \bibinfo{author}{\bibfnamefont{Y.}~\bibnamefont{Liu}}, \bibnamefont{and}
  \bibinfo{author}{\bibfnamefont{A.~M.} \bibnamefont{Goldman}},
  \bibinfo{journal}{Phys. Rev. Lett.} \textbf{\bibinfo{volume}{62}},
  \bibinfo{pages}{2180} (\bibinfo{year}{1989}).

\bibitem[{\citenamefont{Hebard and Paalanen}(1990)}]{HebardPaalanen}
\bibinfo{author}{\bibfnamefont{A.~F.} \bibnamefont{Hebard}} \bibnamefont{and}
  \bibinfo{author}{\bibfnamefont{M.~A.} \bibnamefont{Paalanen}},
  \bibinfo{journal}{Phys. Rev. Lett.} \textbf{\bibinfo{volume}{65}},
  \bibinfo{pages}{927} (\bibinfo{year}{1990}).

\bibitem[{\citenamefont{Steiner et~al.}(2005)\citenamefont{Steiner, Boebinger,
  and Kapitulnik}}]{SteinerKapi}
\bibinfo{author}{\bibfnamefont{M.~A.} \bibnamefont{Steiner}},
  \bibinfo{author}{\bibfnamefont{G.}~\bibnamefont{Boebinger}},
  \bibnamefont{and}
  \bibinfo{author}{\bibfnamefont{A.}~\bibnamefont{Kapitulnik}},
  \bibinfo{journal}{Phys. Rev. Lett.} \textbf{\bibinfo{volume}{94}},
  \bibinfo{eid}{107008} (\bibinfo{year}{2005}).

\bibitem[{\citenamefont{Sambandamurthy
  et~al.}(2004)\citenamefont{Sambandamurthy, Engel, Johansson, and
  Shahar}}]{Shahar1}
\bibinfo{author}{\bibfnamefont{G.}~\bibnamefont{Sambandamurthy}},
  \bibinfo{author}{\bibfnamefont{L.~W.} \bibnamefont{Engel}},
  \bibinfo{author}{\bibfnamefont{A.}~\bibnamefont{Johansson}},
  \bibnamefont{and} \bibinfo{author}{\bibfnamefont{D.}~\bibnamefont{Shahar}},
  \bibinfo{journal}{Phys. Rev. Lett.} \textbf{\bibinfo{volume}{92}},
  \bibinfo{eid}{107005} (\bibinfo{year}{2004}).

\bibitem[{\citenamefont{Frydman et~al.}(2002)\citenamefont{Frydman, Naaman, and
  Dynes}}]{DynesG}
\bibinfo{author}{\bibfnamefont{A.}~\bibnamefont{Frydman}},
  \bibinfo{author}{\bibfnamefont{O.}~\bibnamefont{Naaman}}, \bibnamefont{and}
  \bibinfo{author}{\bibfnamefont{R.~C.} \bibnamefont{Dynes}},
  \bibinfo{journal}{Phys. Rev. B} \textbf{\bibinfo{volume}{66}},
  \bibinfo{pages}{052509} (\bibinfo{year}{2002}).

\bibitem[{\citenamefont{Lau et~al.}(2001)\citenamefont{Lau, Markovic, Bockrath,
  Bezryadin, and Tinkham}}]{Lau}
\bibinfo{author}{\bibfnamefont{C.~N.} \bibnamefont{Lau}},
  \bibinfo{author}{\bibfnamefont{N.}~\bibnamefont{Markovic}},
  \bibinfo{author}{\bibfnamefont{M.}~\bibnamefont{Bockrath}},
  \bibinfo{author}{\bibfnamefont{A.}~\bibnamefont{Bezryadin}},
  \bibnamefont{and} \bibinfo{author}{\bibfnamefont{M.}~\bibnamefont{Tinkham}},
  \bibinfo{journal}{Phys. Rev. Lett.} \textbf{\bibinfo{volume}{87}},
  \bibinfo{pages}{217003} (\bibinfo{year}{2001}).

\bibitem[{\citenamefont{Bezryadin et~al.}(2000)\citenamefont{Bezryadin, Lau,
  and Tinkham}}]{Bezryadin1}
\bibinfo{author}{\bibfnamefont{A.}~\bibnamefont{Bezryadin}},
  \bibinfo{author}{\bibfnamefont{C.~N.} \bibnamefont{Lau}}, \bibnamefont{and}
  \bibinfo{author}{\bibfnamefont{M.}~\bibnamefont{Tinkham}},
  \bibinfo{journal}{Nature} \textbf{\bibinfo{volume}{404}},
  \bibinfo{pages}{971} (\bibinfo{year}{2000}).

\bibitem[{\citenamefont{Rogachev and Bezryadin}(2003)}]{Bezryadin2}
\bibinfo{author}{\bibfnamefont{A.}~\bibnamefont{Rogachev}} \bibnamefont{and}
  \bibinfo{author}{\bibfnamefont{A.}~\bibnamefont{Bezryadin}},
  \bibinfo{journal}{Applied Physics Letters} \textbf{\bibinfo{volume}{83}},
  \bibinfo{pages}{512} (\bibinfo{year}{2003}).

\bibitem[{\citenamefont{Altomare et~al.}(2006)\citenamefont{Altomare, Chang,
  Melloch, Hong, and Tu}}]{Chang}
\bibinfo{author}{\bibfnamefont{F.}~\bibnamefont{Altomare}},
  \bibinfo{author}{\bibfnamefont{A.~M.} \bibnamefont{Chang}},
  \bibinfo{author}{\bibfnamefont{M.~R.} \bibnamefont{Melloch}},
  \bibinfo{author}{\bibfnamefont{Y.}~\bibnamefont{Hong}}, \bibnamefont{and}
  \bibinfo{author}{\bibfnamefont{C.~W.} \bibnamefont{Tu}},
  \bibinfo{journal}{Phys. Rev. Lett.} \textbf{\bibinfo{volume}{97}},
  \bibinfo{eid}{017001} (\bibinfo{year}{2006}).

\bibitem[{\citenamefont{Haviland et~al.}(2000)\citenamefont{Haviland,
  Andersson, and Ågren}}]{Haviland1}
\bibinfo{author}{\bibfnamefont{D.~B.} \bibnamefont{Haviland}},
  \bibinfo{author}{\bibfnamefont{K.}~\bibnamefont{Andersson}},
  \bibnamefont{and} \bibinfo{author}{\bibfnamefont{P.}~\bibnamefont{Ågren}},
  \bibinfo{journal}{J. of Low T. Phys.} \textbf{\bibinfo{volume}{118}},
  \bibinfo{pages}{733} (\bibinfo{year}{2000}).

\bibitem[{\citenamefont{Chow et~al.}(1998)\citenamefont{Chow, Delsing, and
  Haviland}}]{Haviland2}
\bibinfo{author}{\bibfnamefont{E.}~\bibnamefont{Chow}},
  \bibinfo{author}{\bibfnamefont{P.}~\bibnamefont{Delsing}}, \bibnamefont{and}
  \bibinfo{author}{\bibfnamefont{D.~B.} \bibnamefont{Haviland}},
  \bibinfo{journal}{Phys. Rev. Lett.} \textbf{\bibinfo{volume}{81}},
  \bibinfo{pages}{204} (\bibinfo{year}{1998}).

\bibitem[{\citenamefont{Rimberg et~al.}(1997)\citenamefont{Rimberg, Ho, Kurdak,
  Clarke, Campman, and Gossard}}]{Clarke}
\bibinfo{author}{\bibfnamefont{A.~J.} \bibnamefont{Rimberg}},
  \bibinfo{author}{\bibfnamefont{T.~R.} \bibnamefont{Ho}},
  \bibinfo{author}{\bibfnamefont{C.}~\bibnamefont{Kurdak}},
  \bibinfo{author}{\bibfnamefont{J.}~\bibnamefont{Clarke}},
  \bibinfo{author}{\bibfnamefont{K.~L.} \bibnamefont{Campman}},
  \bibnamefont{and} \bibinfo{author}{\bibfnamefont{A.~C.}
  \bibnamefont{Gossard}}, \bibinfo{journal}{Phys. Rev. Lett.}
  \textbf{\bibinfo{volume}{78}}, \bibinfo{pages}{2632} (\bibinfo{year}{1997}).

\bibitem[{\citenamefont{Zhang et~al.}(1989)\citenamefont{Zhang, Hansson, and
  Kivelson}}]{ZHK}
\bibinfo{author}{\bibfnamefont{S.~C.} \bibnamefont{Zhang}},
  \bibinfo{author}{\bibfnamefont{T.~H.} \bibnamefont{Hansson}},
  \bibnamefont{and} \bibinfo{author}{\bibfnamefont{S.}~\bibnamefont{Kivelson}},
  \bibinfo{journal}{Phys. Rev. Lett.} \textbf{\bibinfo{volume}{62}},
  \bibinfo{pages}{82} (\bibinfo{year}{1989}).

\bibitem[{\citenamefont{Oshikawa et~al.}(1997)\citenamefont{Oshikawa, Yamanaka,
  and Affleck}}]{AffleckM}
\bibinfo{author}{\bibfnamefont{M.}~\bibnamefont{Oshikawa}},
  \bibinfo{author}{\bibfnamefont{M.}~\bibnamefont{Yamanaka}}, \bibnamefont{and}
  \bibinfo{author}{\bibfnamefont{I.}~\bibnamefont{Affleck}},
  \bibinfo{journal}{Phys. Rev. Lett.} \textbf{\bibinfo{volume}{78}},
  \bibinfo{pages}{1984} (\bibinfo{year}{1997}).

\bibitem[{\citenamefont{Fisher}(1990)}]{FisherFilms}
\bibinfo{author}{\bibfnamefont{M.~P.~A.} \bibnamefont{Fisher}},
  \bibinfo{journal}{Phys. Rev. Lett.} \textbf{\bibinfo{volume}{65}},
  \bibinfo{pages}{923} (\bibinfo{year}{1990}).

\bibitem[{\citenamefont{Sachdev}(1999)}]{SachdevBook}
\bibinfo{author}{\bibfnamefont{S.}~\bibnamefont{Sachdev}},
  \emph{\bibinfo{title}{Quantum phase transitions}}
  (\bibinfo{publisher}{Cambridge University Press}, \bibinfo{address}{London},
  \bibinfo{year}{1999}).

\bibitem[{\citenamefont{Fisher et~al.}(1989)\citenamefont{Fisher, Weichman,
  Grinstein, and Fisher}}]{FWGF}
\bibinfo{author}{\bibfnamefont{M.~P.~A.} \bibnamefont{Fisher}},
  \bibinfo{author}{\bibfnamefont{P.~B.} \bibnamefont{Weichman}},
  \bibinfo{author}{\bibfnamefont{G.}~\bibnamefont{Grinstein}},
  \bibnamefont{and} \bibinfo{author}{\bibfnamefont{D.~S.}
  \bibnamefont{Fisher}}, \bibinfo{journal}{Phys. Rev. B}
  \textbf{\bibinfo{volume}{40}}, \bibinfo{pages}{546} (\bibinfo{year}{1989}).

\bibitem[{\citenamefont{Mason and Kapitulnik}(2002)}]{Kapi1}
\bibinfo{author}{\bibfnamefont{N.}~\bibnamefont{Mason}} \bibnamefont{and}
  \bibinfo{author}{\bibfnamefont{A.}~\bibnamefont{Kapitulnik}},
  \bibinfo{journal}{Phys. Rev. B} \textbf{\bibinfo{volume}{65}},
  \bibinfo{eid}{220505} (\bibinfo{year}{2002}).

\bibitem[{\citenamefont{Mason and Kapitulnik}(1999)}]{Kapi2}
\bibinfo{author}{\bibfnamefont{N.}~\bibnamefont{Mason}} \bibnamefont{and}
  \bibinfo{author}{\bibfnamefont{A.}~\bibnamefont{Kapitulnik}},
  \bibinfo{journal}{Phys. Rev. Lett.} \textbf{\bibinfo{volume}{82}},
  \bibinfo{pages}{5341} (\bibinfo{year}{1999}).

\bibitem[{\citenamefont{Refael et~al.}(2007)\citenamefont{Refael, Demler, Oreg,
  and Fisher}}]{RDOF2}
\bibinfo{author}{\bibfnamefont{G.}~\bibnamefont{Refael}},
  \bibinfo{author}{\bibfnamefont{E.}~\bibnamefont{Demler}},
  \bibinfo{author}{\bibfnamefont{Y.}~\bibnamefont{Oreg}}, \bibnamefont{and}
  \bibinfo{author}{\bibfnamefont{D.~S.} \bibnamefont{Fisher}},
  \bibinfo{journal}{Phys. Rev. B} \textbf{\bibinfo{volume}{75}},
  \bibinfo{pages}{014522} (\bibinfo{year}{2007}).

\bibitem[{\citenamefont{Michaeli and Finkel'stein}(2006)}]{Fink3}
\bibinfo{author}{\bibfnamefont{K.}~\bibnamefont{Michaeli}} \bibnamefont{and}
  \bibinfo{author}{\bibfnamefont{A.~M.} \bibnamefont{Finkel'stein}},
  \bibinfo{journal}{Phys. Rev. Lett.} \textbf{\bibinfo{volume}{97}},
  \bibinfo{eid}{117004} (\bibinfo{year}{2006}).

\bibitem[{\citenamefont{Vishwanath et~al.}(2004)\citenamefont{Vishwanath,
  Moore, and Senthil}}]{VishMoore}
\bibinfo{author}{\bibfnamefont{A.}~\bibnamefont{Vishwanath}},
  \bibinfo{author}{\bibfnamefont{J.~E.} \bibnamefont{Moore}}, \bibnamefont{and}
  \bibinfo{author}{\bibfnamefont{T.}~\bibnamefont{Senthil}},
  \bibinfo{journal}{Phys. Rev. B} \textbf{\bibinfo{volume}{69}},
  \bibinfo{eid}{054507} (\bibinfo{year}{2004}).

\bibitem[{\citenamefont{Bloch et~al.}(2007)\citenamefont{Bloch, Dalibard, and
  Zwerger}}]{bloch-2007}
\bibinfo{author}{\bibfnamefont{I.}~\bibnamefont{Bloch}},
  \bibinfo{author}{\bibfnamefont{J.}~\bibnamefont{Dalibard}}, \bibnamefont{and}
  \bibinfo{author}{\bibfnamefont{W.}~\bibnamefont{Zwerger}},
  \emph{\bibinfo{title}{Many-body physics with ultracold gases}},
  \bibinfo{howpublished}{arXiv.org:0704.3011} (\bibinfo{year}{2007}).

\bibitem[{\citenamefont{Cramer et~al.}(2004)\citenamefont{Cramer, Eisert, and
  Illuminati}}]{LDA}
\bibinfo{author}{\bibfnamefont{M.}~\bibnamefont{Cramer}},
  \bibinfo{author}{\bibfnamefont{J.}~\bibnamefont{Eisert}}, \bibnamefont{and}
  \bibinfo{author}{\bibfnamefont{F.}~\bibnamefont{Illuminati}},
  \bibinfo{journal}{Phys. Rev. Lett.} \textbf{\bibinfo{volume}{93}},
  \bibinfo{eid}{190405} (\bibinfo{year}{2004}).

\bibitem[{\citenamefont{Albus et~al.}(2003)\citenamefont{Albus, Illuminati, and
  Eisert}}]{albus}
\bibinfo{author}{\bibfnamefont{A.}~\bibnamefont{Albus}},
  \bibinfo{author}{\bibfnamefont{F.}~\bibnamefont{Illuminati}},
  \bibnamefont{and} \bibinfo{author}{\bibfnamefont{J.}~\bibnamefont{Eisert}},
  \bibinfo{journal}{Phys. Rev. A} \textbf{\bibinfo{volume}{68}},
  \bibinfo{pages}{023606} (\bibinfo{year}{2003}).

\bibitem[{\citenamefont{Wang}(2006)}]{Wang}
\bibinfo{author}{\bibfnamefont{D.-W.} \bibnamefont{Wang}},
  \bibinfo{journal}{Phys. Rev. Lett.} \textbf{\bibinfo{volume}{96}},
  \bibinfo{pages}{140404} (\bibinfo{year}{2006}).

\bibitem[{\citenamefont{Titvinidze et~al.}()\citenamefont{Titvinidze, Snoek,
  and Hofstetter}}]{Hofstetter}
\bibinfo{author}{\bibfnamefont{I.}~\bibnamefont{Titvinidze}},
  \bibinfo{author}{\bibfnamefont{M.}~\bibnamefont{Snoek}}, \bibnamefont{and}
  \bibinfo{author}{\bibfnamefont{W.}~\bibnamefont{Hofstetter}},
  \bibinfo{howpublished}{cond-mat/0708.3241}.

\bibitem[{\citenamefont{B\"{u}chler and Blatter}(2004)}]{buechler}
\bibinfo{author}{\bibfnamefont{H.~P.} \bibnamefont{B\"{u}chler}}
  \bibnamefont{and} \bibinfo{author}{\bibfnamefont{G.}~\bibnamefont{Blatter}},
  \bibinfo{journal}{Phys. Rev. A} \textbf{\bibinfo{volume}{69}},
  \bibinfo{eid}{063603} (\bibinfo{year}{2004}).

\bibitem[{\citenamefont{Adhikari and Salasnich}(2007)}]{adhikari}
\bibinfo{author}{\bibfnamefont{S.~K.} \bibnamefont{Adhikari}} \bibnamefont{and}
  \bibinfo{author}{\bibfnamefont{L.}~\bibnamefont{Salasnich}},
  \bibinfo{journal}{Phys. Rev. A} \textbf{\bibinfo{volume}{76}},
  \bibinfo{eid}{023612} (\bibinfo{year}{2007}).

\bibitem[{\citenamefont{Lewenstein et~al.}(2004)\citenamefont{Lewenstein,
  Santos, Baranov, and Fehrmann}}]{LewensteinSantos}
\bibinfo{author}{\bibfnamefont{M.}~\bibnamefont{Lewenstein}},
  \bibinfo{author}{\bibfnamefont{L.}~\bibnamefont{Santos}},
  \bibinfo{author}{\bibfnamefont{M.~A.} \bibnamefont{Baranov}},
  \bibnamefont{and} \bibinfo{author}{\bibfnamefont{H.}~\bibnamefont{Fehrmann}},
  \bibinfo{journal}{Phys. Rev. Lett.} \textbf{\bibinfo{volume}{92}},
  \bibinfo{pages}{050401} (\bibinfo{year}{2004}).

\bibitem[{\citenamefont{Powell et~al.}(2005)\citenamefont{Powell, Sachdev, and
  Buchler}}]{PowellSachdev}
\bibinfo{author}{\bibfnamefont{S.}~\bibnamefont{Powell}},
  \bibinfo{author}{\bibfnamefont{S.}~\bibnamefont{Sachdev}}, \bibnamefont{and}
  \bibinfo{author}{\bibfnamefont{H.~P.} \bibnamefont{Buchler}},
  \bibinfo{journal}{Phys. Rev. B} \textbf{\bibinfo{volume}{72}},
  \bibinfo{pages}{024534} (\bibinfo{year}{2005}).

\bibitem[{\citenamefont{Gunter et~al.}(2006)\citenamefont{Gunter, Stoferle,
  Moritz, Kohl, and Esslinger}}]{Esslinger}
\bibinfo{author}{\bibfnamefont{K.}~\bibnamefont{Gunter}},
  \bibinfo{author}{\bibfnamefont{T.}~\bibnamefont{Stoferle}},
  \bibinfo{author}{\bibfnamefont{H.}~\bibnamefont{Moritz}},
  \bibinfo{author}{\bibfnamefont{M.}~\bibnamefont{Kohl}}, \bibnamefont{and}
  \bibinfo{author}{\bibfnamefont{T.}~\bibnamefont{Esslinger}},
  \bibinfo{journal}{Phys. Rev. Lett.} \textbf{\bibinfo{volume}{96}},
  \bibinfo{pages}{180402} (\bibinfo{year}{2006}).

\bibitem[{\citenamefont{Ospelkaus
  et~al.}(2006{\natexlab{a}})\citenamefont{Ospelkaus, Ospelkaus, Sengstock, and
  Bongs}}]{Bongs}
\bibinfo{author}{\bibfnamefont{C.}~\bibnamefont{Ospelkaus}},
  \bibinfo{author}{\bibfnamefont{S.}~\bibnamefont{Ospelkaus}},
  \bibinfo{author}{\bibfnamefont{K.}~\bibnamefont{Sengstock}},
  \bibnamefont{and} \bibinfo{author}{\bibfnamefont{K.}~\bibnamefont{Bongs}},
  \bibinfo{journal}{Phys. Rev. Lett.} \textbf{\bibinfo{volume}{96}},
  \bibinfo{eid}{020401} (\bibinfo{year}{2006}{\natexlab{a}}).

\bibitem[{\citenamefont{Schreck et~al.}(2001)\citenamefont{Schreck, Khaykovich,
  Corwin, Ferrari, Bourdel, Cubizolles, and Salomon}}]{BECexp}
\bibinfo{author}{\bibfnamefont{F.}~\bibnamefont{Schreck}},
  \bibinfo{author}{\bibfnamefont{L.}~\bibnamefont{Khaykovich}},
  \bibinfo{author}{\bibfnamefont{K.~L.} \bibnamefont{Corwin}},
  \bibinfo{author}{\bibfnamefont{G.}~\bibnamefont{Ferrari}},
  \bibinfo{author}{\bibfnamefont{T.}~\bibnamefont{Bourdel}},
  \bibinfo{author}{\bibfnamefont{J.}~\bibnamefont{Cubizolles}},
  \bibnamefont{and} \bibinfo{author}{\bibfnamefont{C.}~\bibnamefont{Salomon}},
  \bibinfo{journal}{Phys. Rev. Lett.} \textbf{\bibinfo{volume}{87}},
  \bibinfo{pages}{080403} (\bibinfo{year}{2001}).

\bibitem[{\citenamefont{Ospelkaus
  et~al.}(2006{\natexlab{b}})\citenamefont{Ospelkaus, Ospelkaus, Wille, Succo,
  Ernst, Sengstock, and Bongs}}]{Bongs2}
\bibinfo{author}{\bibfnamefont{S.}~\bibnamefont{Ospelkaus}},
  \bibinfo{author}{\bibfnamefont{C.}~\bibnamefont{Ospelkaus}},
  \bibinfo{author}{\bibfnamefont{O.}~\bibnamefont{Wille}},
  \bibinfo{author}{\bibfnamefont{M.}~\bibnamefont{Succo}},
  \bibinfo{author}{\bibfnamefont{P.}~\bibnamefont{Ernst}},
  \bibinfo{author}{\bibfnamefont{K.}~\bibnamefont{Sengstock}},
  \bibnamefont{and} \bibinfo{author}{\bibfnamefont{K.}~\bibnamefont{Bongs}},
  \bibinfo{journal}{Phys. Rev. Lett.} \textbf{\bibinfo{volume}{96}},
  \bibinfo{eid}{180403} (\bibinfo{year}{2006}{\natexlab{b}}).

\bibitem[{\citenamefont{Modugno et~al.}(2003)\citenamefont{Modugno, Ferlaino,
  Riboli, Roati, Modugno, and Inguscio}}]{inguscio}
\bibinfo{author}{\bibfnamefont{M.}~\bibnamefont{Modugno}},
  \bibinfo{author}{\bibfnamefont{F.}~\bibnamefont{Ferlaino}},
  \bibinfo{author}{\bibfnamefont{F.}~\bibnamefont{Riboli}},
  \bibinfo{author}{\bibfnamefont{G.}~\bibnamefont{Roati}},
  \bibinfo{author}{\bibfnamefont{G.}~\bibnamefont{Modugno}}, \bibnamefont{and}
  \bibinfo{author}{\bibfnamefont{M.}~\bibnamefont{Inguscio}},
  \bibinfo{journal}{Phys. Rev. A} \textbf{\bibinfo{volume}{68}},
  \bibinfo{pages}{043626} (\bibinfo{year}{2003}).

\bibitem[{\citenamefont{Ferlaino et~al.}(2004)\citenamefont{Ferlaino,
  de~Mirandes, Roati, Modugno, and Inguscio}}]{Inguscio-exp}
\bibinfo{author}{\bibfnamefont{F.}~\bibnamefont{Ferlaino}},
  \bibinfo{author}{\bibfnamefont{E.}~\bibnamefont{de~Mirandes}},
  \bibinfo{author}{\bibfnamefont{G.}~\bibnamefont{Roati}},
  \bibinfo{author}{\bibfnamefont{G.}~\bibnamefont{Modugno}}, \bibnamefont{and}
  \bibinfo{author}{\bibfnamefont{M.}~\bibnamefont{Inguscio}},
  \bibinfo{journal}{Phys. Rev. Lett.} \textbf{\bibinfo{volume}{92}},
  \bibinfo{pages}{140405} (\bibinfo{year}{2004}).

\bibitem[{\citenamefont{Hadzibabic et~al.}(2002)\citenamefont{Hadzibabic, Stan,
  Dieckmann, Gupta, Zwierlein, G\"orlitz, and Ketterle}}]{HadziFB}
\bibinfo{author}{\bibfnamefont{Z.}~\bibnamefont{Hadzibabic}},
  \bibinfo{author}{\bibfnamefont{C.~A.} \bibnamefont{Stan}},
  \bibinfo{author}{\bibfnamefont{K.}~\bibnamefont{Dieckmann}},
  \bibinfo{author}{\bibfnamefont{S.}~\bibnamefont{Gupta}},
  \bibinfo{author}{\bibfnamefont{M.~W.} \bibnamefont{Zwierlein}},
  \bibinfo{author}{\bibfnamefont{A.}~\bibnamefont{G\"orlitz}},
  \bibnamefont{and} \bibinfo{author}{\bibfnamefont{W.}~\bibnamefont{Ketterle}},
  \bibinfo{journal}{Phys. Rev. Lett.} \textbf{\bibinfo{volume}{88}},
  \bibinfo{pages}{160401} (\bibinfo{year}{2002}).

\bibitem[{\citenamefont{Roethel and Pelster}()}]{hydro}
\bibinfo{author}{\bibfnamefont{S.}~\bibnamefont{Roethel}} \bibnamefont{and}
  \bibinfo{author}{\bibfnamefont{A.}~\bibnamefont{Pelster}},
  \bibinfo{howpublished}{cond-mat/0703220}.

\bibitem[{\citenamefont{Pollet et~al.}()\citenamefont{Pollet, Kollath,
  Schollw\"{o}ck, and Troyer}}]{Kollath}
\bibinfo{author}{\bibfnamefont{L.}~\bibnamefont{Pollet}},
  \bibinfo{author}{\bibfnamefont{C.}~\bibnamefont{Kollath}},
  \bibinfo{author}{\bibfnamefont{U.}~\bibnamefont{Schollw\"{o}ck}},
  \bibnamefont{and} \bibinfo{author}{\bibfnamefont{M.}~\bibnamefont{Troyer}},
  \bibinfo{howpublished}{cond-mat/0609604}.

\bibitem[{\citenamefont{Wen}(2004)}]{Wen}
\bibinfo{author}{\bibfnamefont{X.-G.} \bibnamefont{Wen}},
  \emph{\bibinfo{title}{Quantum Field Theory of Many-body Systems}}
  (\bibinfo{publisher}{Oxford University Press}, \bibinfo{year}{2004}).

\bibitem[{\citenamefont{Mahan}(1981)}]{Mahan}
\bibinfo{author}{\bibfnamefont{G.~D.} \bibnamefont{Mahan}},
  \emph{\bibinfo{title}{Many-Particle Physics}} (\bibinfo{publisher}{AP},
  \bibinfo{year}{1981}).

\bibitem[{\citenamefont{Anderson}(1967)}]{AndersonOC}
\bibinfo{author}{\bibfnamefont{P.~W.} \bibnamefont{Anderson}},
  \bibinfo{journal}{Phys. Rev. Lett.} \textbf{\bibinfo{volume}{18}},
  \bibinfo{pages}{1049} (\bibinfo{year}{1967}).

\bibitem[{\citenamefont{Yang}(2007)}]{yang-2007}
\bibinfo{author}{\bibfnamefont{K.}~\bibnamefont{Yang}},
  \emph{\bibinfo{title}{Superfluid-insulator transition and fermion pairing in
  bose-fermi mixtures}}, \bibinfo{howpublished}{arXiv.org:0707.4189}
  (\bibinfo{year}{2007}).

\bibitem[{\citenamefont{Mering and Fleischhauer}(2007)}]{Fleischhauer}
\bibinfo{author}{\bibfnamefont{A.}~\bibnamefont{Mering}} \bibnamefont{and}
  \bibinfo{author}{\bibfnamefont{M.}~\bibnamefont{Fleischhauer}},
  \emph{\bibinfo{title}{The one-dimensional bose-fermi-hubbard model in the
  heavy-fermion limit}}, \bibinfo{howpublished}{arXiv.org:0709.2386}
  (\bibinfo{year}{2007}).

\bibitem[{\citenamefont{Mathey et~al.}(2004)\citenamefont{Mathey, Wang,
  Hofstetter, Lukin, and Demler}}]{MatheyWang}
\bibinfo{author}{\bibfnamefont{L.}~\bibnamefont{Mathey}},
  \bibinfo{author}{\bibfnamefont{D.-W.} \bibnamefont{Wang}},
  \bibinfo{author}{\bibfnamefont{W.}~\bibnamefont{Hofstetter}},
  \bibinfo{author}{\bibfnamefont{M.~D.} \bibnamefont{Lukin}}, \bibnamefont{and}
  \bibinfo{author}{\bibfnamefont{E.}~\bibnamefont{Demler}},
  \bibinfo{journal}{Phys. Rev. Lett.} \textbf{\bibinfo{volume}{93}},
  \bibinfo{eid}{120404} (pages~\bibinfo{numpages}{4}) (\bibinfo{year}{2004}).

\bibitem[{\citenamefont{Sengupta and Pryadko}(2005)}]{sengupta-2005}
\bibinfo{author}{\bibfnamefont{P.}~\bibnamefont{Sengupta}} \bibnamefont{and}
  \bibinfo{author}{\bibfnamefont{L.~P.} \bibnamefont{Pryadko}},
  \emph{\bibinfo{title}{Quantum degenerate bose-fermi mixtures on 1-d optical
  lattices}}, \bibinfo{howpublished}{cond-mat/0512241} (\bibinfo{year}{2005}).

\bibitem[{\citenamefont{Altman and Auerbach}(1998)}]{Altman}
\bibinfo{author}{\bibfnamefont{E.}~\bibnamefont{Altman}} \bibnamefont{and}
  \bibinfo{author}{\bibfnamefont{A.}~\bibnamefont{Auerbach}},
  \bibinfo{journal}{Phys. Rev. Lett.} \textbf{\bibinfo{volume}{81}},
  \bibinfo{pages}{4484} (\bibinfo{year}{1998}).

\bibitem[{\citenamefont{Nozi\'eres and De~Dominicis}(1969)}]{Nozieres}
\bibinfo{author}{\bibfnamefont{P.}~\bibnamefont{Nozi\'eres}} \bibnamefont{and}
  \bibinfo{author}{\bibfnamefont{C.~T.} \bibnamefont{De~Dominicis}},
  \bibinfo{journal}{Phys. Rev.} \textbf{\bibinfo{volume}{178}},
  \bibinfo{pages}{1097} (\bibinfo{year}{1969}).

\bibitem[{\citenamefont{Yuval and Anderson}(1970)}]{AndersonYuval}
\bibinfo{author}{\bibfnamefont{G.}~\bibnamefont{Yuval}} \bibnamefont{and}
  \bibinfo{author}{\bibfnamefont{P.~W.} \bibnamefont{Anderson}},
  \bibinfo{journal}{Phys. Rev. B} \textbf{\bibinfo{volume}{1}},
  \bibinfo{pages}{1522} (\bibinfo{year}{1970}).

\bibitem[{\citenamefont{Caldeira and Leggett}(1981)}]{C-L1}
\bibinfo{author}{\bibfnamefont{A.~O.} \bibnamefont{Caldeira}} \bibnamefont{and}
  \bibinfo{author}{\bibfnamefont{A.~J.} \bibnamefont{Leggett}},
  \bibinfo{journal}{Phys. Rev. Lett.} \textbf{\bibinfo{volume}{46}},
  \bibinfo{pages}{211} (\bibinfo{year}{1981}).

\bibitem[{\citenamefont{Caldeira and Leggett}(1983)}]{C-L2}
\bibinfo{author}{\bibfnamefont{A.~O.} \bibnamefont{Caldeira}} \bibnamefont{and}
  \bibinfo{author}{\bibfnamefont{A.~J.} \bibnamefont{Leggett}},
  \bibinfo{journal}{Ann. Phys.} \textbf{\bibinfo{volume}{\bf 149}},
  \bibinfo{pages}{374} (\bibinfo{year}{1983}).

\bibitem[{foo({\natexlab{a}})}]{footnoteX}
\bibinfo{howpublished}{But since $\gamma$ can not exceed 1, the qualitative
  nature of the SF-INS transition is always the same, and never becomes a
  Caldeira-Leggett type transition.}

\bibitem[{\citenamefont{Schmid}(1983)}]{Schmid}
\bibinfo{author}{\bibfnamefont{A.}~\bibnamefont{Schmid}},
  \bibinfo{journal}{Phys. Rev. Lett.} \textbf{\bibinfo{volume}{51}},
  \bibinfo{pages}{1506} (\bibinfo{year}{1983}).

\bibitem[{\citenamefont{Chakravarty}(1982)}]{Chakravarty}
\bibinfo{author}{\bibfnamefont{S.}~\bibnamefont{Chakravarty}},
  \bibinfo{journal}{Phys. Rev. Lett.} \textbf{\bibinfo{volume}{49}},
  \bibinfo{pages}{681} (\bibinfo{year}{1982}).

\bibitem[{\citenamefont{Ambegaokar et~al.}(1982)\citenamefont{Ambegaokar,
  Eckern, and Sch\"on}}]{AES}
\bibinfo{author}{\bibfnamefont{V.}~\bibnamefont{Ambegaokar}},
  \bibinfo{author}{\bibfnamefont{U.}~\bibnamefont{Eckern}}, \bibnamefont{and}
  \bibinfo{author}{\bibfnamefont{G.}~\bibnamefont{Sch\"on}},
  \bibinfo{journal}{Phys. Rev. Lett.} \textbf{\bibinfo{volume}{48}},
  \bibinfo{pages}{1745} (\bibinfo{year}{1982}).

\bibitem[{foo({\natexlab{b}})}]{footnote1}
\bibinfo{howpublished}{Note, also, that the mean-field is strictly only valid
  at $d\ge 3$, and long-range phase ordering may only appear at $d\ge 2$ by the
  Mermin-Wagner theorem.}

\bibitem[{\citenamefont{Greiner et~al.}(2002)\citenamefont{Greiner, Mandel,
  H\"ansch, and Bloch}}]{Bloch-revival}
\bibinfo{author}{\bibfnamefont{M.}~\bibnamefont{Greiner}},
  \bibinfo{author}{\bibfnamefont{O.}~\bibnamefont{Mandel}},
  \bibinfo{author}{\bibfnamefont{T.~W.} \bibnamefont{H\"ansch}},
  \bibnamefont{and} \bibinfo{author}{\bibfnamefont{I.}~\bibnamefont{Bloch}},
  \bibinfo{journal}{Nature} \textbf{\bibinfo{volume}{419}}, \bibinfo{pages}{51}
  (\bibinfo{year}{2002}).

\end{thebibliography}

\end{document}